\documentclass[journal]{IEEEtran}

\pdfoutput=1

\usepackage{cite}
\usepackage{fixltx2e}

\usepackage{amsfonts}
\usepackage[cmex10]{amsmath}
\usepackage{amssymb,graphics,times,bm,bbm,color}

\usepackage{hyperref}
\usepackage[normalem]{ulem}

\usepackage{amsthm} 

\ifCLASSINFOpdf
   \usepackage[pdflatex]{graphicx}
\else
   \usepackage[dvips]{graphicx}
\fi

\usepackage[font=footnotesize]{subfig}

\renewcommand{\baselinestretch}{1.}
\newcommand{\ignore}[1]{}

\newcommand{\T}{\mathcal{T}}
\newcommand{\V}{\mathcal{V}}
\newcommand{\J}{\mathcal{J}}
\newcommand{\A}{\mathcal{A}}
\newcommand{\AtA}{{\mathcal{A}}{\times}{\mathcal{A}}}

\newtheorem{mytheorem}{Theorem}
\newtheorem{mylemma}{Lemma}

\newtheorem{mydef}{Definition}
\newtheorem{mycond}{Condition}
\newtheorem{myconda}{Condition}
\newcommand{\beq} {\begin{equation}}
\newcommand{\eeq} {\end{equation}}
\newcommand{\bea} {\begin{eqnarray}}
\newcommand{\eea} {\end{eqnarray}}
\newcommand{\bes} {\begin{subequations}}
\newcommand{\ees} {\end{subequations}}

\begin{document}

\title{Quantum adiabatic machine learning}
\author{Kristen L. Pudenz, Daniel A. Lidar, ~\IEEEmembership{Senior~Member,~IEEE}%
\thanks{The authors are grateful to the Lockheed Martin Corporation for financial support under the URI program. KP is also supported by the NSF under a graduate research fellowship. DAL acknowledges support from the NASA Ames Research Center.}
\thanks{K. L. Pudenz is with the Department of Electrical Engineering and the Center
for Quantum Information Science and Technology, University of Southern California,
Los Angeles, CA 90089 USA (e-mail: pudenz@usc.edu).}%
\thanks{D. A. Lidar is with the Departments of Electrical Engineering, Chemistry,
Physics and the Center for Quantum Information Science and Technology, University
of Southern California, Los Angeles CA 90089 USA (e-mail: lidar@usc.edu).}
}

%


\maketitle

\begin{abstract}
We develop an approach to machine learning and anomaly detection via quantum adiabatic evolution. In the training phase we identify an optimal set of weak classifiers, to form a single strong classifier. In the testing phase we adiabatically evolve one or more strong classifiers on a superposition of inputs in order to find certain anomalous elements in the classification space. Both the training and testing phases are executed via quantum adiabatic evolution. We apply and illustrate this approach in detail to the problem of software verification and validation.
\end{abstract}


\IEEEpeerreviewmaketitle

\section{Introduction}

\IEEEPARstart{M}{achine} learning is a field of computational research 
with broad applications, 
ranging from image processing to analysis of complex systems such as the stock 
market. There is abundant literature concerning learning theory in the classical 
domain, addressing speed and accuracy of the learning process for different classes 
of concepts \cite{Vapnik:book}. Groundwork for machine learning using quantum computers has also been 
laid, showing that quantum machine learning, while requiring as much input 
information as classical machine learning, may be faster and is capable of handling 
concepts beyond the reach of any classical learner \cite{Servedio,Aimeur}.

We consider the machine learning problem of binary classification, assigning a data 
vector to one of two groups based on criteria derived from a set of training examples 
provided to the algorithm beforehand. The learning method we use is boosting, whereby 
multiple \textit{weak classifiers} are combined to create a \textit{strong classifier} 
formula that is more accurate than any of its components alone \cite{Meir,Freund}. This method 
can be applied to any problem where the separation of two groups of data is required, 
whether it is distinguishing two species of plants based on their measurements or picking 
out the letter \textquotedblleft a\textquotedblright~ from all other letters of the 
alphabet when it is scanned. Our approach to classification is based on recent efforts 
in boosting using adiabatic quantum optimization (AQO) which showed advantages over 
classical boosting in the sparsity of the classifiers achieved and their accuracy (for 
certain problems)\cite{Neven1,Neven3}.

As a natural outgrowth of the classification problem, we also formulate a scheme for 
anomaly detection using quantum computation. Anomaly detection has myriad uses, some 
examples of which are detection of insider trading, finding faults in mechanical 
systems, and highlighting changes in time-lapsed satellite imagery \cite{Chandola:2009}. 
Specifically, we pursue the verification and validation (V\&V) of classical software, with 
programming errors as the anomalies to be detected. This is one of the more 
challenging potential applications of quantum anomaly detection, because programs 
are large, complex, and highly irregular in their structure. However, it is also an 
important and currently intractable problem for which even small gains are likely to 
yield benefits for the software development and testing community.

The complexity of the V\&V problem is easily understood by considering the number of 
operations necessary for an exhaustive test of a piece of software. Covering 
every possible set of inputs that could be given to the software requires a 
number of tests that is exponential in the number of input variables, 
notwithstanding the complexity of each individual test \cite{EWD}. Although 
exhaustive testing is infeasible due to its difficulty, the cost of this 
infeasibility is large - in 2002, NIST estimated that tens of billions of dollars 
were lost due to inadequate testing \cite{Tassey}.

The subject of how to best implement software testing given limited resources 
has been widely studied. Within this field, efforts focused on combinatorial 
testing have found considerable success and will be relevant to our new approach. 
Combinatorial testing focuses on using the test attempts available to test all 
combinations of up to a small number, $t$, of variables, with the idea that errors 
are usually caused by the interaction of only a few parameters \cite{Bryce,Kuhn}. 
This approach has found considerable success \cite{Grindal,Cohen}, with scaling that 
is logarithmic in $n$, the number of software parameters, and exponential in $t$. 

Currently, the use of formal methods in the coding and verification phases of software 
development is the only way to guarantee absolute correctness of software without implementing 
exhaustive testing. However, formal methods,  
are also expensive and time-consuming to implement. Model checking, a method of software 
analysis which aims to ensure the validity of all reachable program states, solves 
$n$-bit satisfiability problems (which are NP-complete), with $n$ as a function of the 
number of reachable states of the program \cite{DSilva}. Theorem proving, where a program is developed 
alongside a proof of its own correctness, requires repeated interaction and correction 
from the developer as the proof is formed, with the intermediate machine-provable lemmas 
checked with a satisfiability solver \cite{Weber}.

We propose a new approach to verification and validation of software which makes 
use of quantum information processing. The approach consists of a quantum learning step and a quantum
testing step. In the learning step, our strategy uses quantum optimization to learn 
the characteristics of the program being tested and the specification it is being 
tested to. This learning technique is known as quantum boosting and has been previously 
applied to other problems, in particular image recognition \cite{Neven,Neven1,Neven2,Bian}. 
Boosting consists 
of building up a formula to accurately sort inputs into one of two groups by combining 
simple rules that sort less accurately, and in its classical forms has been frequently 
addressed in the machine learning literature \cite{Meir,Schapire,Freund}.

The testing step is novel, and involves turning the classifying formulas 
generated by the learning step into a function that generates a lower energy the more 
likely its input is to represent a software error. This function is translated into the 
problem Hamiltonian of an adiabatic quantum computation (AQC). The AQC allows all 
potential software errors (indeed, as we will see, all possible operations of the software) 
to be examined in quantum-parallel, returning only the best candidates for errors which 
correspond to the lowest values of the classification function.

Both the learning and testing steps make use of AQC. An 
adiabatic quantum algorithm encodes the desired result in the ground state of some problem  
Hamiltonian. The computation is then performed by initializing a physical system in the 
easily prepared ground state of a simpler Hamiltonian, then slowly changing the control 
parameters of the system so the system undergoes an adiabatic evolution to the ground 
state of the difficult-to-solve problem Hamiltonian \cite{Farhi1,Farhi2}. The adiabatic 
model of quantum 
computation is known to be universal and equivalent to the circuit model with a 
polynomial conversion overhead \cite{Aharonov:04,Mizel}. While it is not known at this time how to make AQC fault tolerant, several error correction and prevention protocols have been proposed for AQC \cite{Jordan:05,Lidar:AQC-DD}, and it is known to exhibit a certain degree of natural robustness \cite{Childs,SarandyLidar:05}.

In this article, Section 
\ref{formal_section}
will begin by establishing the framework 
through which the quantum V\&V problem is attacked, 
and by defining 
the programming errors we seek to eliminate. As we proceed with the development 
of a method for V\&V using quantum resources, Section \ref{sec_train} will establish {an implementation of} the learning step as an adiabatic quantum algorithm. {We develop conditions for ideal boosting and an alternate quantum learning algorithm in Section \ref{conditions}.} The testing 
step will be detailed in Section \ref{sec_use}. We present simulated results of 
the learning step on a sample problem in Section \ref{simulation_section}, and finish with 
our conclusions and suggestions for future work in Section \ref{conclusions}.

\section{Formalization}
\label{formal_section}

In this section we formalize the problem of software error detection by first introducing the relevant vector spaces and then giving a criterion for the occurrence of an error.

\subsection{Input and output spaces}
\label{spaces_section}

Consider an \textquotedblleft ideal\textquotedblright\ software program $\hat{P}$,
where by ideal we mean the correct program which a perfect programmer would
have written. Instead we are faced with the real life implementation of
$\hat{P}$, which we denote by $P$ and refer to as the ``implemented program." Suppose we wish to verify the
operation of $P$ relative to $\hat{P}$. All programs have input and output
spaces $\mathcal{V}_{\rm in}$ and $\mathcal{V}_{\rm out}$, such that
\begin{equation}
P:\mathcal{V}_{\rm in}\mapsto\mathcal{V}_{\rm out}.
\end{equation}
Without loss of generality we can think of these spaces as being spaces of
binary strings. This is so because the input to any program is always specified
within some finite machine precision, and the output is again given within
finite machine precision (not necessarily the same as the input precision).
Further, since we are only interested in inputs
and outputs which take a finite time to generate (or \textquotedblleft write
down\textquotedblright), without loss of generality we can set upper limits on the lengths of allowed
input and output strings.   Within these constraints we can move to a binary
representation for both input and output spaces, and take $N_{\rm in}$ as the
maximum number of bits required to specify any input, and $N_{\rm out}$ as the
maximum number of bits required to specify any output. Thus we can identify
the input and output spaces as binary string spaces%
\begin{equation}
\mathcal{V}_{\rm in}\cong\{0,1\}^{N_{\rm in}},\quad\mathcal{V}_{\rm out}\cong\{0,1\}^{N_{\rm out}}.
\end{equation}

It will be convenient to concatenate elements of the input and output spaces
into single objects. Thus, consider binary vectors $\vec{x}=(\vec{x}_{\rm in}
,\vec{x}_{\rm out})$, where $\vec{x}_{\rm out}=P(\vec{x}_{\rm in})$, consisting of program
input-output pairs:\
\begin{equation}
\vec{x}\in\{0,1\}^{N_{\rm in}}\times\{0,1\}^{N_{\rm out}}=\{0,1\}^{N_{\rm in}+N_{\rm out}
}\equiv\mathcal{V}.
\end{equation}

\subsection{Recognizing software errors}
\label{error_section}

\subsubsection{Validity domain and range}
We shall assume without loss of generality that the input spaces of the ideal and implemented programs are identical. This can always be ensured by extending the ideal program so that it is well defined for all elements of $\mathcal{V}_{\rm in}$. Thus, while in general not all elements of $\mathcal{V}_{\rm in}$ have to be allowed inputs into
$\hat{P}$ (for example, an input vector that is out of range for the ideal program), one can always reserve some fixed value for such inputs (e.g., the largest vector in ${\cal V}_{\rm out}$) and trivially mark them as errors. 
The ideal program $\hat{P}$ is thus a map from the input space to
the space $\mathcal{R}_{\rm out}$ of correct outputs:
\begin{equation}
\hat{P}:\mathcal{V}_{\rm in}\mapsto\mathcal{R}_{\rm out} \subseteq {\cal V}_{\rm out}. 
\label{Phat}
\end{equation}
More specifically, $\hat{P}$ computes an output string $\hat{x}_{\rm out}$ for
every input string $\vec{x}_{\rm in}$, i.e., we can write $\hat{x}_{\rm out}=\hat
{P}(\vec{x}_{\rm in})$. 
Of course this map can be many-to-one (non-injective and surjective), but not
one-to-many (multi-valued).\footnote{Random number generation may appear to be a
counterexample, as it is multi-valued, but only over different calls
to the random-number generator.}
 The implemented program 
$P$ should ideally compute the exact
same function. In reality it may not. 
With this in mind, the simplest way to identify a software error is to find an input vector $\vec{x}_{\rm in}$ such that
\begin{equation}
\| \hat{P}(\vec{x}_{\rm in})-P(\vec{x}_{\rm in})\| \neq0.
\label{error1}
\end{equation}
in some appropriate norm. This is clearly a sufficient condition for an error, since the implemented program must agree with the ideal program on all inputs. However, for our purposes a more general approach will prove to be more suitable.

\subsubsection{Specification and implementation sets}
\label{sets_section}

A direct way to think about the existence of errors in a software program 
is to 
consider two ordered sets within the space of input-output pairs, $\mathcal{V}$. 
These are the set of ordered, correct input-output pairs $\hat{S}$ according to 
the program specification $\hat{P}$, and the set of input-output pairs $S$ 
implemented by the real program $P$. 
We call $\hat{S}$ the ``specification set'' and $S$ the ``implementation set''. The program under test is correct when 
\begin{equation}
\hat{S}=S .
\label{noerror}
\end{equation}
That is, in a correct program, the 
specification
set of correct input-output pairs is exactly 
the set that is implemented in code.

As stated,
(\ref{noerror}) is impractical since it requires knowledge of the 
complete structure of the intended input and output spaces. Instead, we can also use the specification and implementation sets to give a correctness criterion for a given input-output pair:
\begin{mydef}
A vector $\vec{x} \in \mathcal{V}$ is erroneous and implemented if 
\begin{equation}
\vec{x} \notin \hat{S} \quad \& \quad \vec{x} \in S .
\label{eq:impl-err}
\end{equation}
\label{def:impl-err}
\end{mydef}
Input-output vectors satisfying ~\eqref{eq:impl-err} are the manifestation of software errors (``bugs") and their identification is the main problem we are concerned with here. 
Conversely, we have
\begin{mydef}
A vector $\vec{x} \in \mathcal{V}$ is correct and implemented if 
\begin{equation}
\vec{x} \in \hat{S} \quad \& \quad \vec{x} \in S .
\label{eq:impl-cor}
\end{equation}
\label{def:impl-cor}
\end{mydef}
Input-output vectors satisfying ~\eqref{eq:impl-cor} belong to the ``don't-worry'' class. 
The two other possibilities belong to the ``don't-care'' class:
\begin{mydef}
A vector $\vec{x} \in \mathcal{V}$ is correct and unimplemented if 
\begin{equation}
\vec{x} \in \hat{S} \quad \& \quad \vec{x} \notin S .
\label{eq:unimpl-cor}
\end{equation}
\label{def:unimpl-cor}
\end{mydef}
\begin{mydef}
A vector $\vec{x} \in \mathcal{V}$ is erroneous and unimplemented if 
\begin{equation}
\vec{x} \notin \hat{S} \quad \& \quad \vec{x} \notin S .
\label{eq:unimpl-err}
\end{equation}
\label{def:unimpl-err}
\end{mydef}

\begin{figure}
\centering
\includegraphics[width=0.45\textwidth]{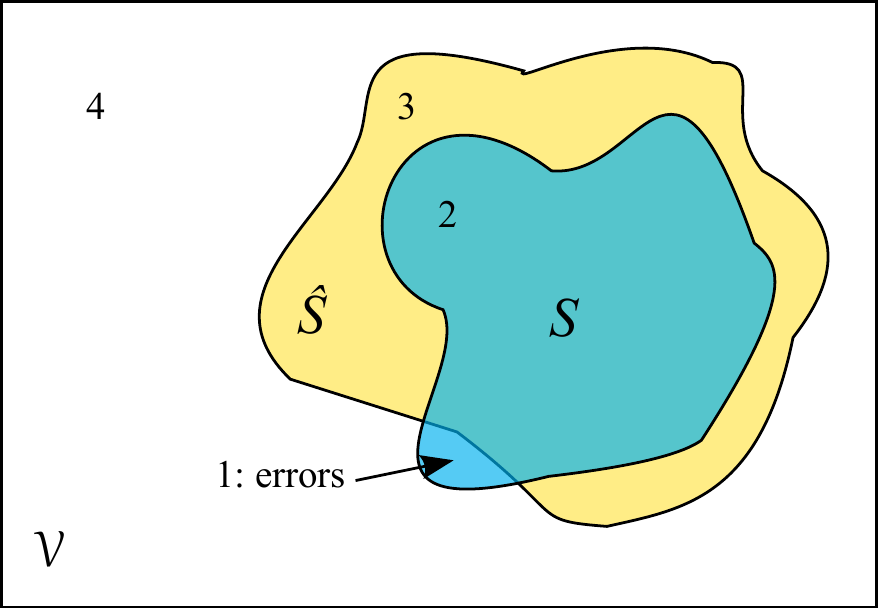}
\caption{Schematic vector space representation showing regions of vectors satisfying the four definitions. Region 1, of erroneous but implemented vectors,  is the location of errors. Regions 2, 3, and 4 represent vectors which are correct and implemented, correct and unimplemented, and erroneous and unimplemented, respectively.}
\label{vec_space}
\end{figure}

A representation of the locations of vectors satisfying the four definitions for a sample vector space can be found in Fig. \ref{vec_space}. Our focus will be on the erroneous vectors of Definition \ref{def:impl-err}.

Note that Eq.~\eqref{error1} implies that the vector is erroneous and implemented, i.e., Definition~\ref{def:impl-err}. Indeed, let $\vec{x}_{\rm out} = P(\vec{x}_{\rm in})$, i.e., $\vec{x} = (\vec{x}_{\rm in},\vec{x}_{\rm out}) \in S$, but assume that $\vec{x}_{\rm out} \neq \hat{x}_{\rm out}$ where $\hat{x}_{\rm out} = \hat{P}(\vec{x}_{\rm in})$. Then $\vec{x} \notin \hat{S}$, since $\vec{x}_{\rm in}$ pairs up with $\hat{x}_{\rm out}$ in $\hat{S}$.
Conversely, Definition~\ref{def:impl-err} implies Eq.~\eqref{error1}. To see this, assume that $\vec{x} = (\vec{x}_{\rm in},\vec{x}_{\rm out}) \in S$ but $\vec{x} = (\vec{x}_{\rm in},\vec{x}_{\rm out}) \notin \hat{S}$. This must mean that $\vec{x}_{\rm out} \neq \hat{x}_{\rm out}$, again because $\vec{x}_{\rm in}$ pairs up with $\hat{x}_{\rm out}$ in $\hat{S}$. Thus Eq.~\eqref{error1} is in fact equivalent to Definition~\ref{def:impl-err}, but does not capture the other three possibilities captured by Definitions \ref{def:impl-cor}-\ref{def:unimpl-err}.

Definitions~\ref{def:impl-err}-\ref{def:unimpl-err} will play a central role in our approach to quantum V\&V. 

\subsubsection{Generalizations}
Note that it may well be advantageous in practice to consider a more general
setup, where instead of studying only the map from the input to the output
space, we introduce intermediate maps which track intermediate program states.
This can significantly improve our error classification accuracy.\footnote{One
important consideration is that, as we shall see below, for practical reasons we
may only be able to track errors at the level of one-bit errors and
correlations between bit-pairs. Such limited tracking can be alleviated to
some extent by using intermediate spaces, where higher order correlations
between bits appearing at the level of the output space may not yet
have had time
to develop.} Formally, this would mean that Eq.~(\ref{Phat}) is replaced by
\begin{equation}
\hat{P}:\mathcal{V}_{\rm in}\mapsto\mathcal{I}_{1}\mapsto\cdots\mapsto
\mathcal{I}_{J}\mapsto\mathcal{R}_{\rm out},
\label{int}
\end{equation}
where $\{\mathcal{I}_{j}\}_{j=1}^{J}$ are intermediate spaces. 
However, we shall not consider 
this more refined approach
in this work.

As a final general comment, we reiterate that a solution of the problem we have defined has implications beyond V\&V. Namely, Definitions~\ref{def:impl-err}-\ref{def:unimpl-err} capture a broad class of anomaly (or outlier) detection problems \cite{Chandola:2009}. From this perspective the approach we detail in what follows can be  described as ``quantum anomaly detection,'' and could be pursued in any application which requires the batch processing of a large data space to find a few anomalous elements.

\section{Training a quantum software error classifier}
\label{sec_train}

In this section we discuss how to identify whether a given set of 
input-output pairs is erroneous or correct, and implemented or unimplemented, as per Definitions~\ref{def:impl-err}-\ref{def:unimpl-err}.
To this end we shall require so-called {\it weak classifiers}, a
{\it strong classifier}, a methodology to
efficiently train the strong classifier,
and a way to
efficiently apply the trained strong classifier on all possible input-output pairs. 
Both the training step and the application step will potentially benefit from a quantum speedup.

\subsection{Weak classifiers}

Consider a class of functions which map from the input-output space to the
reals:
\begin{equation}
h_{i}:\mathcal{V}\mapsto\mathbb{R}.
\end{equation}
We call these functions ``weak classifiers" or ``feature detectors," where
$i\in\{1,...,N\}$ enumerates the features. These are some predetermined useful
aggregate characteristics of the program $P$ which we can measure, such as
total memory, or CPU time average \cite{Stehle}. Note that $N$ will turn out
to be the number of qubits we shall require in our quantum approach.

We can now formally associate a weak classification with each vector in the input-output space. 
\begin{mydef}Weak classification of $\vec{x}\in \mathcal{V}$.\\
Weakly classified correct (WCC): a vector $\vec{x}$ is WCC if $h_{i}(\vec{x})>0$.\\
Weakly classified erroneous (WCE): a vector $\vec{x}$ is WCE if $h_{i}(\vec{x})<0$.
\label{def:WC}
\end{mydef}

Clearly, there is an advantage to finding \textquotedblleft
smart\textquotedblright\ weak classifiers, so as to minimize $N$. This 
can be done by invoking heuristics, 
or via 
a systematic approach 
such as one we present below.

For each input-output pair $\vec{x}$ we have a
vector $\vec{h}(\vec{x})=\left(  h_{1}(\vec{x}),...,h_{N}(\vec{x})\right)
\in\mathbb{R}^{N}$. Such vectors can be used to construct geometric
representations of the learning problem, e.g., a convex hull encompassing the
weak classifier vectors of clustered correct input-output pairs. Such a
computational geometry approach was pursued in \cite{Stehle}.

We assume that we can construct a \textquotedblleft training
set\textquotedblright\ 
\beq\mathcal{T}\equiv\{\vec{x}_{s},y_{s}\}_{s=1}^{S},
\label{trainset}
\eeq
where each $\vec{x}_{s}\in \mathcal{V}$ is an input-output pair and
$y_{s}=y(\vec{x}_{s})=+1$ iff $\vec{x}_{s}$ is 
correct (whether implemented or not, i.e., $\vec{x}_s \in \hat{S}$)
while $y_{s}=-1$ iff $\vec
{x}_{s}$ 
is erroneous (again, implemented or not, i.e., $\vec{x}_s \notin \hat{S}$).
Thus, the training set represents the
ideal program $\hat{P}$, i.e., we assume that the training set can be
completely trusted. Note that Eq.~(\ref{Phat}) presents us with an easy method
for including erroneous input pairs, by deliberately 
misrepresenting the action of $\hat{P}$ on some given input, e.g., by setting 
$\vec{x}_{\rm out} \notin {\cal R}_{\rm out}(\hat{P})$. This is
similar to the idea of 
performing V\&V by
building invariants into a program  \cite{LeTraon}. 

We are free to normalize each weak classifier so that $h_{i}\in\lbrack
-1/N,1/N]$ (the reason for this will become clear below). Given Definition~\ref{def:WC} we choose the sign of each weak classifier so that
$h_{i}(\vec{x}_{s})<0$ for all erroneous training data, while $h_{i}(\vec{x}_{s})>0$ for all correct training data. Each point $\vec{h}(\vec{x}_{s})\in\lbrack-1/N,1/N]^{N}$ (a hypercube) has associated with it a label $y_{s}$
which indicates whether the point is correct or erroneous. The convex hull approach to V\&V \cite{Stehle} assumes that correct training points $\vec{h}(\vec{x}_{s})$ cluster. Such an assumption is not required in our approach.

\subsection{Strong classifier}

We would like to combine all the weak classifiers into a single
\textquotedblleft strong classifier\textquotedblright\ which, given an
input-output pair, will determine that pair's correctness or erroneousness. The
problem is that we do not know in advance how to rank the weak classifiers by
relative importance. We can formally solve this problem by associating a
weight $w_{i}\in\mathbb{R}$ with each weak classifier $h_{i}$. The problem
then becomes how to find the optimal set of weights, given the training set.

The process of creating a high-performance strong classifier from many less 
accurate weak classifiers is known as boosting in the machine learning literature. 
Boosting is a known method for enhancing to arbitrary levels the performance of 
known sets of classifiers that exhibit weak learnability for a problem, i.e., 
they are accurate on more than half of the training set \cite{Schapire,Mannor}. 
The most efficient 
method to combine weak classifiers into a strong classifier of a given accuracy 
is an open question, and there are many competing algorithms available for this 
purpose \cite{Kotsiantis,Yu}. Issues commonly considered in the development of such algorithms 
include identification of the data features that are relevant to the classification 
problem at hand \cite{Zhang,Cheng} and whether or not provisions need to be taken to avoid 
overfitting to the training set (causing poor performance on the general problem 
space) \cite{Breiman,Blumer}. We use an approach inspired by recent quantum boosting 
results on image recognition \cite{Neven,Neven1,Neven2,Bian}. This 
approach has been shown to outperform classical boosting algorithms in terms of accuracy (but not speed) on selected 
problems, and has the advantage of being implementable on existing quantum 
optimization hardware 
\cite{Biamonte,Choi:08,Karimi,Harris}.

Since we shall map the $w_{i}$ to qubits we use binary weights $w_{i} \in\{0,1\}$.
It should be straightforward to generalize our approach to
a higher resolution version of real-valued
$w_{i}$ using multiple qubits per weight.

Let $\vec{w}=(w_{1},...,w_{N})\in\{0,1\}^{N}$, and let
\begin{equation}
R_{\vec{w}}(\vec{x})\equiv\vec{w}\cdot\vec{h}(\vec{x})=\sum_{i=1}^{N} w_{i}h_{i}(\vec{x})\in\lbrack-1,1]. 
\label{R}
\end{equation}
This range is a direct result of the normalization $h_{i}\in\lbrack-1/N,1/N]$
introduced above.

We now define the weight-dependent \textquotedblleft strong
classifier\textquotedblright\ 
\begin{equation}
Q_{\vec{w}}(\vec{x})\equiv\mathrm{sign}\left[  R_{\vec{w}}(\vec{x})\right],
\label{Q}
\end{equation}
and use it as follows:
\begin{mydef}
Strong classification of $\vec{x}\in \mathcal{V}$.\\
Strongly classified correct (SCC): a vector $\vec{x}$ is SCC if $Q_{\vec{w}}(\vec{x})=+1$.\\
Strongly classified erroneous (SCE): a vector $\vec{x}$ is SCE if $Q_{\vec{w}}(\vec{x})=-1$.
\label{def:SC}
\end{mydef}

There is a fundamental difference between the \textquotedblleft
opinions\textquotedblright\ of the strong classifier, as expressed in Definition~\ref{def:SC}, and the actual
erroneousness/correctness of a given input-output pair. The strong classifier
associates an erroneous/correct label with a given input-output pair according to a
weighted average of the weak classifiers. This opinion may or may not be
correct. For the training set we actually know whether a given input-output
pair is erroneous or correct. This presents us with an opportunity to compare the
strong classifier to the training data. Namely, if $y_{s}Q_{\vec{w}}(\vec
{x}_{s})=-1$ then $Q_{\vec{w}}(\vec{x}_{s})$ and $y_{s}$ have opposite sign,
i.e., disagree, which means that $Q_{\vec{w}}(\vec{x}_{s})$ mistakenly
classified $\vec{x}_{s}$ as a correct input-output pair while in fact it was
erroneous, or vice versa. On the other hand, if $y_{s}Q_{\vec{w}}(\vec{x}_{s})=+1$
then $Q_{\vec{w}}(\vec{x}_{s})$ and $y_{s}$ agree, which means that
$Q_{\vec{w}}(\vec{x}_{s})$ is correct. Formally,
\bes
\label{eq:class}
\bea
y_{s}Q_{\vec{w}}(\vec{x}_{s})=+1 &\Longleftrightarrow& \begin{cases}
(\vec{x}_{s} \text{ is SCC})=\text{true or}\\
(\vec{x}_{s} \text{ is SCE})=\text{true}
\end{cases}\\
y_{s}Q_{\vec{w}}(\vec{x}_{s})=-1 &\Longleftrightarrow& \begin{cases}
(\vec{x}_{s} \text{ is SCC})=\text{false or}\\
(\vec{x}_{s} \text{ is SCE})=\text{false}
\end{cases}
\eea
\ees
The higher the number of true instances is relative to the number of false instances, the better the strong classifier performance over the training set. The challenge is, of course, to construct a strong classifier that performs well also beyond the training set. To do so we must first solve the problem of finding the optimal set of binary weights $\vec{w}$.

\subsection{The formal weight optimization problem}

Let $\mathbf{H}\left[  z\right]  $ denote the Heaviside step function, i.e.,
$\mathbf{H}\left[  z\right]  =0$ if $z<0$ and $\mathbf{H}\left[  z\right]  =1$
if $z>0$. Thus $\mathbf{H}\left[  -y_{s}Q_{\vec{w}}(\vec{x}_{s})\right]  =1$
if the classification of $\vec{x}_{s}$ is wrong, but $\mathbf{H}\left[
-y_{s}Q_{\vec{w}}(\vec{x}_{s})\right]  =0$ if the classification of $\vec
{x}_{s}$ is correct. In this manner $\mathbf{H}\left[  -y_{s}Q_{\vec{w}}%
(\vec{x}_{s})\right]  $ assigns a penalty of one unit for each incorrectly
classified input-output pair.

Consider%
\begin{equation}
L(\vec{w})\equiv\sum_{s=1}^{S}\mathbf{H}\left[  -y_{s}Q_{\vec{w}}%
(x_{s})\right]  . 
\label{L}
\end{equation}
This counts the total number of incorrect classifications. Therefore
minimization of $L(\vec{w})$ for a given training set $\{\vec{x}_{s}%
,y_{s}\}_{s=1}^{S}$ will yield the optimal set of weights $\vec{w}%
^{\rm opt}=\{w_{i}^{\rm opt}\}_{i=1}^{N}$.

However, it is important
not to overtrain the classifier. Overtraining means that the strong classifier has poor generalization performance, i.e., it does not classify accurately outside of the training set \cite{Blumer,Cheng:07}. To prevent overtraining we can add a penalty proportional to
the Hamming weight of $\vec{w}$, i.e., to the number of non-zero weights
$\| \vec{w}\| _{0}=\sum_{i=1}^{N}w_{i}$. In this manner an optimal balance is sought between the accuracy of the strong classifier and the number of weak classifiers comprising the strong classifier. The formal weight
optimization problem is then to solve
\begin{equation}
\vec{w}^{\prime opt}=\arg\min_{\vec{w}}\left[  L(w)+\lambda\| \vec
{w}\| _{0}\right]  , 
\label{w-opt}
\end{equation}
where $\lambda>0$ can be tuned to decide the relative importance of the penalty.

\subsection{Relaxed weight optimization problem}
\label{sec:relaxed}

Unfortunately, the formulation of ~(\ref{w-opt}) is unsuitable for
adiabatic quantum computation 
because of its discrete nature. In particular, the evaluation of the Heaviside
function is not amenable to a straightforward implementation in AQC. Therefore, following \cite{Neven1}, we now
relax it by introducing a quadratic error measure, which will be implementable
in AQC.

Let $\vec{y}=(y_{1},...,y_{S})\in\{-1,1\}^{S}$ and $\vec{R}_{\vec{w}}%
=(R_{\vec{w}}(\vec{x}_{1}),...,R_{\vec{w}}(\vec{x}_{S}))\in\lbrack-1,1]^{S}$.
The vector $\vec{y}$ is the ordered label set of correct/erroneous input-output
pairs. The components $R_{\vec{w}}(\vec{x})$ of the vector $\vec{R}%
_{\vec{w}}$ already appeared in the strong classifier 
(\ref{Q}). There we were interested only in their signs and in Eq.~\eqref{eq:class} we observed that if
$y_{s}R_{\vec{w}}(\vec{x}_{s})<0$ then $\vec{x}_{s}$ was incorrectly
classified, while if $y_{s}R_{\vec{w}}(\vec{x}_{s})>0$ then $\vec{x}_{s}$ was
correctly classified.

We can consider a relaxation of the formal optimization problem (\ref{w-opt})
by replacing the counting of incorrect classifications 
by a sum of the values of $y_{s}R_{\vec{w}}(\vec{x}_{s})$ over the training
set. This makes sense since we have normalized the weak classifiers so that
$R_{\vec{w}}(\vec{x})\in\lbrack-1,1]$, while each label $y_{s}\in\{-1,1\}$, so
that all the terms $y_{s}R_{\vec{w}}(\vec{x}_{s})$ are in principle equally
important. In other words, the inner product $\vec{y}\cdot\vec{R}_{\vec{w}%
}=\sum_{s=1}^{S}y_{s}R_{\vec{w}}(\vec{x}_{s})$ is also a measure of the
success of the classification, and maximizing it (making $\vec{y}$ and
$\vec{R}_{\vec{w}}$ as parallel as possible) should result in a good training set.

Equivalently,
we can
consider the distance between the vectors $\vec{y}$ and $\vec{R}_{\vec{w}}$
and minimize it by finding the optimal weight vector $\vec{w}^{\rm opt}$, in
general different from that in Eq.~(\ref{w-opt}). Namely, consider the
Euclidean distance%
\begin{align}
\delta(\vec{w})  &  =\| \vec{y}-\vec{R}_{\vec{w}}\| ^{2}=\sum
_{s=1}^{S}\left\vert y_{s}-\sum_{i=1}^{N}w_{i}h_{i}(x_{s})\right\vert
^{2}\nonumber\\
&  =\| \vec{y}\| ^{2}+\sum_{i,j=1}^{N}C_{ij}^{\prime}%
w_{i}w_{j}-2\sum_{i=1}^{N}C_{iy}^{\prime}w_{i},
\end{align}
where $\vec{h}_{i}=(h_{i}(x_{1}),...,h_{i}(x_{S}))\in\lbrack-1/N,1/N]^{S}$ and
where%
\begin{align}
C_{ij}^{\prime}&=\vec{h}_{i}\cdot\vec{h}_{j}=\sum_{s=1}^{S}h_{i}(x_{s}%
)h_{j}(x_{s}),\\
\quad C_{iy}^{\prime}&=\vec{h}_{i}\cdot\vec{y}=\sum_{s=1}%
^{S}h_{i}(x_{s})y_{s}%
\end{align}
can be thought of as correlation functions. Note that they are
symmetric:$\ C_{ij}^{\prime}=C_{ji}^{\prime}$ and $C_{iy}^{\prime}%
=C_{yi}^{\prime}$. The term $\| \vec{y}\| ^{2}=S$ is a
constant offset so can be dropped from the minimization.

If we wish to introduce a sparsity penalty as above, we can do so again, and thus ask
for the optimal weight in the following sense:%
\begin{align}
\vec{w}^{\rm opt}  &  =\arg\min_{\vec{w}}\left[  \delta(\vec{w})+\lambda^{\prime
}\| \vec{w}\| _{0}\right] \nonumber\\
&  =\arg\min_{\vec{w}}\left[  \sum_{i,j=1}^{N}C_{ij}^{\prime}w_{i}w_{j}%
+2\sum_{i=1}^{N}(\lambda-C_{iy}^{\prime})w_{i}\right]  , 
\label{QUBO}%
\end{align}
where $\lambda^{\prime}=2\lambda$.

\subsection{From QUBO\ to the Ising Hamiltonian}
\label{QUBO2Ising}

Equation (\ref{QUBO}) is a quadratic binary optimization (QUBO) problem \cite{Neven}. One
more step is needed before we can map it to qubits, since we need to work with
optimization variables whose range is $\{-1,1\}$, not $\{0,1\}$. Define new
variables $q_{i}=2(w_{i}-1/2)\in\{-1,1\}$. In terms of these new variables the
minimization problem is%
\begin{align}
\vec{q}^{\rm opt}  &  =\arg\min_{\vec{q}}\left[  \frac{1}{4}\sum_{i,j=1}^{N}%
C_{ij}^{\prime}(q_{i}+1)(q_{j}+1)\right.\nonumber\\
&\qquad \left. +\sum_{i=1}^{N}(\lambda-C_{iy}^{\prime})(q_{i}+1)\right] \nonumber\\
&  =\arg\min_{\vec{q}}\left[  \sum_{i,j=1}^{N}C_{ij}q_{i}q_{j}+\sum_{i=1}%
^{N}(\lambda-C_{iy})q_{i}\right]  , \label{q_opt}%
\end{align}
where in the second line we dropped the constant terms $\frac{1}{4}%
\sum_{i,j=1}^{N}C_{ij}^{\prime}$ and $\sum_{i=1}^{N}(\lambda-C_{iy}^{\prime}%
)$, used the symmetry of $C_{ij}^{\prime}$ for $\sum_{i=1}^{N}q_{i}\sum
_{j=1}^{N}C_{ij}^{\prime}=\sum_{i,j=1}^{N}C_{ij}^{\prime}q_{j}$, and where we
defined
\begin{equation}
C_{ij}=\frac{1}{4}C_{ij}^{\prime},\quad C_{iy}=C_{iy}^{\prime}-\frac{1}{2}%
\sum_{j=1}^{N}C_{ij}^{\prime}. \label{Cij}%
\end{equation}

Thus, the final AQC Hamiltonian for the quantum weight-learning problem is
\begin{equation}
H_{F}=\sum_{i,j=1}^{N}C_{ij}Z_{i}Z_{j}+\sum_{i=1}^{N}(\lambda-C_{iy})Z_{i},
\label{Hf}
\end{equation}
where $Z_{i}$ is the Pauli spin-matrix $\sigma_{z}$ acting on the $i$th qubit.
This represents Ising spin-spin interactions with coupling matrix $C_{ij}$,
and an inhomogeneous magnetic field $\lambda-C_{iy}$ acting on each spin. Note
how $H_{F}$ encodes the training data $\{h_{i}(x_{s}),y_{s}\}_{i,s}$ via the
coupling matrix $C_{ij}=\frac{1}{4}\sum_{s=1}^{S}h_{i}(x_{s})h_{j}(x_{s})$ and
the local magnetic field $C_{iy}=\sum_{s=1}^{S}h_{i}(x_{s})y_{s}-\frac{1}%
{2}\sum_{s=1}^{S}h_{i}(x_{s})\sum_{j=1}^{N}h_{j}(x_{s})$. Thus, in order to generate
$H_{F}$ one must first calculate the training data using the chosen set of
weak classifiers.

In this final form [Eq.~(\ref{Hf})], involving only one and two-qubit $Z_{i}$ terms, 
the problem is now suitable for implementation on devices such as D-Wave's 
adiabatic quantum optimization processor \cite{Bian,Choi:08}. 

In Section~\ref{sec:alternate} we shall formulate an alternative weight optimization problem, based on a methodology we develop in Section~\ref{conditions} for pairing weak classifiers to guarantee the correctness of the strong classifier.

\subsection{Adiabatic quantum computation}
The adiabatic
quantum algorithm implements the time-dependent interpolation%
\begin{equation}
H(t)=s(t)H_{I}+[1-s(t)] H_{F}, 
\label{interp}
\end{equation}
where $H_{I}$ is a Hamiltonian which does not commute with $H_{F}$ and should have a ground state (lowest-energy eigenvector) that is easily reachable, such as 
\beq
H_{I}=\mathbb{I}-\sum_{i=1}^{N}X_{i}
\label{H_I}
\eeq 
where $\mathbb{I}$ is the identity operator and $X_{i}$ is the Pauli $\sigma_{x}$ acting on
the $i$th qubit \cite{Farhi1,Farhi2}. The interpolation function $s(t)$ satisfies the boundary
conditions $s(0)  =1$, $s(T)  =0$, where $T$ is the
final time. Provided the evolution is sufficiently slow (in a manner we shall quantify momentarily), the adiabatic theorem
guarantees that the final state $|\psi(T)\rangle$ reached by the algorithm is,
with high probability, the one that minimizes the energy of $H_{F}$
\cite{Teufel:book,Jansen:06,LidarAT:08}. This means that, for $H_F$ chosen as in Eq.~\eqref{Hf}, it finds as a
ground state the
optimal weights vector $\vec{q}^{\rm opt}$ as defined in ~(\ref{q_opt}). These weights can then be ``read
off" by measuring the final states of each of the $N$ qubits: $|\psi(T)\rangle
=|q_{1}^{\rm opt},...,q_{N}^{\rm opt}\rangle=|\vec{q}^{\rm opt}\rangle$.

It should be noted that while the number of weak classifiers that can be selected from using this algorithm may appear to be limited by the number of qubits available for processing, this is not in fact the case. By performing multiple rounds of optimization, each time filling in the spaces left by classifiers that were assigned weight $0$ in the previous round, an optimized group of $N$ weak classifiers can be assembled. If the performance of the strong classifier is unsatisfactory with $N$ weak classifiers, multiple groups of $N$ found in this manner may be used together.


The scaling of the computation time $t_F$ with the number of qubits (or weak classifiers, in our case), $N$, is determined by the inverse of the minimal ground state energy gap of $H(t)$. There are many variants of the adiabatic theorem, differing mostly in assumptions about boundary conditions and differentiability of $H(t)$. Most variants state that, provided 
\beq
t_F\gtrsim  \frac{\|\dot{H}\|^\alpha}{\epsilon \Delta^{\alpha+1}},
\eeq
then 
\beq
|\langle \psi(t_F)|\phi(t_F)\rangle | \gtrsim 1-\epsilon^\beta .
\label{overlap}
\eeq 
The left-hand side of Eq.~\eqref{overlap} is the fidelity of the \emph{actual} state $|\psi(t_F)\rangle$ obtained under quantum evolution subject to $H(t)$ with respect to the \emph{desired} final ground state $|\phi(t_F)\rangle$. More precisely, $|\psi(t)\rangle$ is the solution of the time-dependent Schr\"{o}dinger equation $\partial |\psi(t)\rangle/\partial t = -iH(t)|\psi(t)\rangle$ (in $\hbar\equiv 1$ units), and $|\phi(t)\rangle$ is the instantaneous ground state of $H(t)$, i.e., the solution of $H(t)|\phi(t)\rangle = E_0(t)|\phi(t)\rangle$, where $E_0(t)$ is the instantaneous ground state energy [the smallest eigenvalue of $H(t)$]. The parameter $\epsilon$, $0\leq \epsilon \leq 1$, measures the quality of the overlap between $|\psi(t_F)\rangle$ and $|\phi(t_F)\rangle$, $\dot{H}$ is the derivative with respect to the dimensionless time $t/t_F$, $\Delta$ is the minimum energy gap between the ground state $|\phi(t)\rangle$ and the first excited state of $H(t)$ (i.e., the difference between the two smallest equal-time eigenvalues of $H(t)$, for $t\in[0,t_F]$), The values of the integers $\alpha$ and $\beta$ depend on the assumptions made about the boundary conditions and differentiability of $H(t)$ \cite{Teufel:book,Jansen:06,LidarAT:08}; typically $\alpha \in \{0,1,2\}$, while $\beta$ can be tuned between $1$ and arbitrarily large values, depending on boundary conditions determining the smoothness of $H(t)$ (see, e.g., Theorem 1 in Ref.~\cite{LidarAT:08}). The crucial point is that the gap $\Delta$ depends on $N$, typically shrinking as $N$ grows, while the numerator $\|\dot{H}\|$ typically has a mild $N$-dependence (bounded in most cases by a function growing as $N^2$ \cite{LidarAT:08}). Consequently a problem has an efficient, polynomial time solution under AQC if $\Delta$ scales $1/{\rm poly}(N)$. However, note that an inverse exponential gap dependence on $N$ can still result in a speedup, as is the case, e.g., in the adiabatic implementation of Grover's search problem \cite{Roland:02,RPL:10}, where the speedup relative to classical computation is quadratic.

As for the problem we are concerned with here, finding the ground state of $H_F$ as prescribed in Eq.~\eqref{Hf} in order to find the optimal weight
set for the (relaxed version of the) problem of training a software
error-classifier, it is not known whether it is amenable to a quantum speedup. A study of the gap dependence of our Hamiltonian $H(t)$ on $N$, which is beyond the scope of the present work, will help to determine whether such a speedup is to be expected also in the problem at hand. A related image processing problem has been shown numerically to require fewer weak classifiers than in comparable classical algorithms, which gives the strong classifier a lower Vapnik-Chernovenkis dimension and therefore a lower generalization error \cite{Neven2,Neven3}. Quantum boosting applied to a different task, $30$-dimensional clustering, demonstrated increasingly better accuracy as the overlap between the two clusters grew than that exhibited by the classical AdaBoost algorithm \cite{Neven1}. More generally, numerical simulations of quantum adiabatic implementations of related hard optimization problems (such as Exact Cover) have 
shown promising scaling results for $N$ values of up to 128 \cite{Farhi2,young:170503,Karimi}. We shall thus proceed here with the requisite cautious optimism.

\section{Achievable strong classifier accuracy}
\label{conditions}

We shall show in this section that it is theoretically possible to construct a perfect, 100\% accurate majority-vote strong 
classifier from a set of weak classifiers that are more than 50\% accurate - 
if those weak classifiers relate to each other in exactly the right way. Our construction in this section is analytical and exact; we shall specify a set of conditions weak classifiers should satisfy for perfect accuracy of the strong classifier they comprise. We shall also show how to construct an imperfect strong classifier, with bounded error probability, by a relaxation of the conditions we shall impose on the weak classifiers. We expect the quantum algorithm to find a close approximation to this result.

Consider 
a strong classifier with a general binary weight vector $\vec{w}\in\{0,1\}^N$, as defined in Eq.~\eqref{R}.
Our approach will be to show that the strong classifier in Eq.~(\ref{R}) is completely accurate if a set of three 
conditions is met. The conditions work by using pairs of weak classifiers which 
both classify some $\vec{x}$ correctly and which disagree for all other $\vec{x}$.
An accurate strong classifier can be constructed by covering the entire space 
$\mathcal{V}$ with the correctly classifying portions of such weak classifier pairs. 

To start, every vector $\vec{x}\in \mathcal{V}$ has a correct classification, as determined by the specification set: 
\bes
\bea
\vec{x}\in \hat{S} &\Longleftrightarrow& y(\vec{x})=+1,\\
\vec{x}\notin \hat{S} &\Longleftrightarrow& y(\vec{x})=-1
\eea
\ees
A strong classifier is perfect if
\beq
{Q_{\vec{w}}(\vec{x})=y(\vec{x}) \quad \forall \vec{x}\in\V.}
\label{eq:perfect-strong}
\eeq
The weak classifiers either agree or disagree with this correct classification. We define the correctness value of a weak classifier for a given input $\vec{x}$:
\begin{equation}
c_i(\vec{x})=h_i(\vec{x})y(\vec{x})=\begin{cases}
+1 & h_i(\vec{x})=y(\vec{x})\\
-1 & h_i(\vec{x})\neq y(\vec{x})
\end{cases}
\end{equation}
Thus, similarly to the strong classifier case [Eq.~\eqref{eq:class}] we have, formally,
\bes
\label{eq:classW}
\bea
c_i(\vec{x})=+1 &\Longleftrightarrow& \begin{cases}
(\vec{x} \text{ is WCC})=\text{true or}\\
(\vec{x} \text{ is WCE})=\text{true}
\end{cases}\\
c_i(\vec{x})=-1 &\Longleftrightarrow& \begin{cases}
(\vec{x} \text{ is WCC})=\text{false or}\\
(\vec{x} \text{ is WCE})=\text{false}
\end{cases}
\eea
\ees
{where WCC and WCE stand for weakly classified correct and weakly classified erroneous, respectively (Definition~\ref{def:WC}).}

A given input-output vector $\vec{x}$ receives either a true or false vote from each weak classifier comprising the strong classifier. Let us denote the index set of the weak classifiers comprising a given strong classifier by $\mathcal{I}$. If the majority of the votes given by the weak classifiers in $\mathcal{I}$ {are} true then the vector receives a strong classification that is true. Let us loosely denote by $\vec{w}\in\mathcal{I}$ the set of weak classifiers whose indices all belong to $\mathcal{I}$. Thus
\beq
\sum_{i\in\mathcal{I}}c_i(\vec{x}) > 0 \Longrightarrow Q_{\vec{w}}(\vec{x})=y(\vec{x}) \text{ if } \vec{w}\in\mathcal{I}.
\label{eq:perfect-weak}
\eeq
It follows from Eq.~\eqref{eq:perfect-strong} that if we can find a set of weak classifiers for which $\sum_{i\in\mathcal{I}}c_i(\vec{x}) > 0$ for all input-output vectors $\vec{x}$, then the corresponding strong classifier is perfect. This is what we shall set out to do in the next subsection.

\subsection{Conditions for complete classification accuracy}
First, we limit our working set to those weak classifiers with greater than 50\% 
accuracy. This is a prerequisite for the feasibility of the other conditions. To ensure that at least half the initial dictionary of weak 
classifiers is more than 50\% accurate, we include each potential weak classifier 
in the dictionary, as well as its opposite. The opposite classifier follows 
the same rule as its counterpart, but makes the opposite binary decision every time,
making each right where the other is wrong and ensuring that at least one of them 
will have 50\% or greater accuracy. Condition 1, therefore, defines the set 
$\mathcal{A}$, 
\beq
\mathcal{A}\subseteq\mathcal{D}\equiv\{1,...,N\},
\eeq
of sufficiently accurate weak classifiers, where $\mathcal{D}$ is the set of all possible values of the index $i$ of weak classifiers in Eq.~(\ref{R}). 

\begin{mycond} 
\label{cond1}
For an input-output vector $\vec{x}\in \mathcal{V}$ selected uniformly at random
\beq
\mathcal{A}=\{i:P[c_i(\vec{x})=1]>1/2\}.
\label{eq:cond1}
\eeq
\end{mycond}
$P[\omega]$ denotes the probability of event $\omega$. We use a probabilistic formulation for our conditions since we imagine the input-output space $\mathcal{V}$ to be very large and accessed by random sampling. 

Conditions \ref{cond2} and \ref{cond3} (or \ref{cond3a}) specify the index set 
\beq
\mathcal{J}\subseteq\mathcal{A}\times{\mathcal{A}},
\eeq
labeling pairs of weak classifiers which will make up the final strong classifier. Condition \ref{cond2} groups the weak classifiers into pairs which classify the minimal number of vectors 
$\vec{x}$ correctly at the same time and give opposite classifications on all other 
vectors. Condition \ref{cond3} completes the specification of the index set $\mathcal{J}$: it states that the subsets of vectors $\vec{x}$ that are classified 
correctly by the classifier pairs in $\mathcal{J}$ must cover the entire space $\mathcal{V}$.

\begin{mycond}
\label{cond2}
If $(j,j') \in \mathcal{J}$ then
\begin{equation}
\begin{split}
& P\left[\left(c_{j}(\vec{x})=1\right){\cap} \left(c_{j'}(\vec{x})=1
\right)\right] \\
& = P\left[c_{j}(\vec{x})=1\right] + P\left[c_{j'}(\vec{x})
=1\right] - 1 
\end{split}
\label{cond_2}
\end{equation}
for an input-output vector $\vec{x}\in \V$ selected uniformly at random.
\end{mycond}

\begin{figure}
\centering
\includegraphics[width=0.45\textwidth]{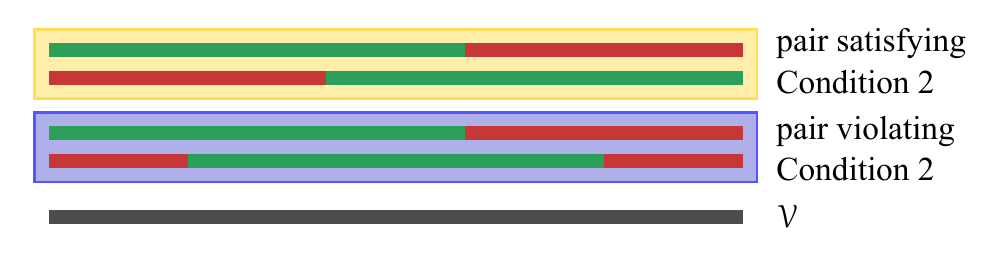}
\caption{Illustration of Condition~\ref{cond2}. Two pairs of classifiers showing regions of correct (green) and incorrect (red) classification along a line representing a lexicographical ordering of all vectors within $\mathcal{V}$. The top pair, compliant with Condition 2, provides two correct classifications for the minimum possible number of vectors, voting once correctly and once incorrectly on all other vectors. The bottom pair, violating Condition 2, provides two correct votes for more vectors than does the top pair, but also undesirably provides two incorrect votes for some vectors; this is why paired weak classifiers must coincide in their classifications on as few vectors as possible.}
\label{cond_2_fig}
\end{figure}

This condition has the following simple interpretation, illustrated in Fig.~\ref{cond_2_fig}. Suppose the entire input-output space $\V$ is sorted lexicographically (e.g., according to the binary values of the vectors $\vec{x}\in\V$) so that the $j$th weak classifier is correct on all first $N_j$ vectors but erroneous on the rest, while the $j'$th weak classifier is correct on all last $N_{j'}$ vectors but erroneous on the rest. Thus the fraction of correctly classified vectors by the $j$th classifier is {$(1-\eta_j) = N_j/|\V|$}, the fraction of correctly classified vectors by the $j'$th classifier is {$(1-\eta_{j'}) = N_{j'}/|\V|$}, and they overlap on a fraction of {$1-\eta_j-\eta_{j'}$}  vectors {(all vectors minus each classifier's fraction of incorrectly classified vectors)}, as illustrated in the top part of Fig.~\ref{cond_2_fig}. By ``pushing classifier $j'$ to the left'', as illustrated in the bottom part of Fig.~\ref{cond_2_fig}, the overlap grows and is no longer minimal. This is what is expressed by Eq.~\eqref{cond_2}.

\begin{figure}
\centering
\includegraphics[width=0.45\textwidth]{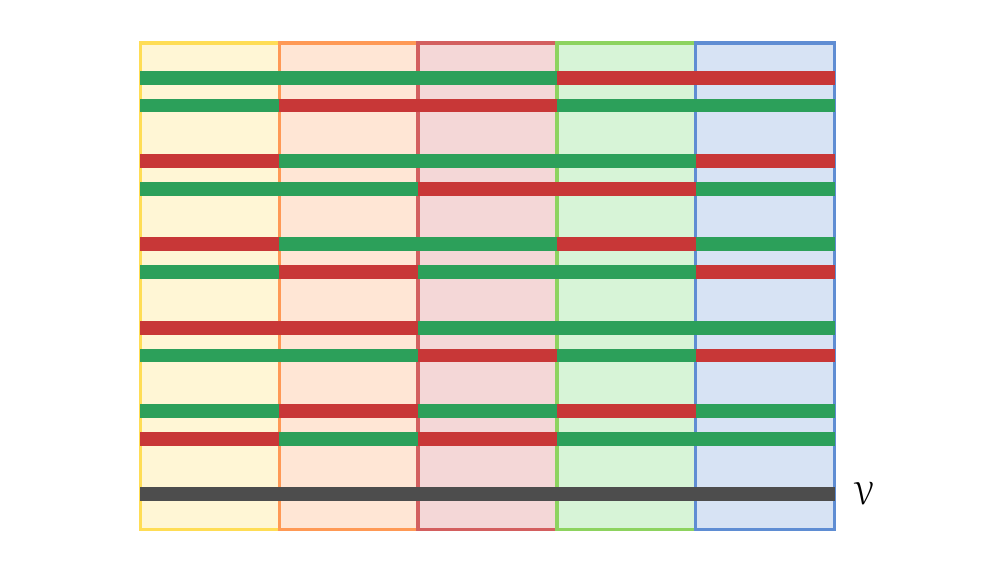}
\caption{Illustration of Condition~\ref{cond3} without the subtracted term. Five pairs of 60\% accurate weak classifiers combine to form a completely accurate majority-vote strong classifier. Moving from top to bottom through the pairs and from left to right along the vectors in the classification space, each pair of weak classifiers provides two correct votes for 20\% of the vector space and neutral votes otherwise. This means that the majority vote is correct for the entire space because no two pairs vote correctly at once.}
\label{cond_3_fig}
\end{figure}

\begin{figure}
\centering
\includegraphics[width=0.45\textwidth]{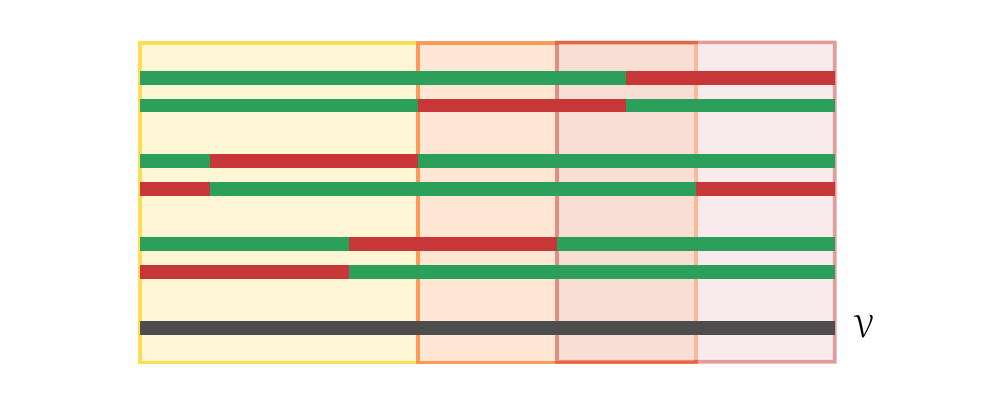}
\caption{Illustration of Condition~\ref{cond3} with the subtracted term. Three pairs of 70\% accurate weak classifiers combined to form a completely accurate majority-vote strong classifier. In this case, each pair votes twice correctly on 40\% of the vector space, which makes it necessary for the correct portions of the second and third pairs from the top to overlap. Because they only overlap by the minimum amount necessary, $\V$ as a whole is still covered by a correct majority vote.}
\label{cond_3_fig_2}
\end{figure}

Condition \ref{cond2} considers only one pair of weak classifiers at a time, which does not suffice to cover all of $\V$. Consider a set of weak classifier pairs each satisfying Condition \ref{cond2} which, together, do cover all of $\V$. Such a set would satisfy $\sum_{(j,j') \in \mathcal{J}}P\left[\left(c_{j}(\vec{x})=1\right) \cap \left(c_{j'}
(\vec{x})=1\right)\right] = 1$ for a randomly chosen $\vec{x}\in\V$. This is illustrated in Fig.~\ref{cond_3_fig}. However, it is also possible for two or more pairs to overlap, a situation we would like to avoid as much as possible, i.e., we shall impose minimal overlap similarly to Condition \ref{cond2}. Thus we arrive at:

\begin{mycond}
\label{cond3}
\begin{equation}
\begin{split}
& \sum_{(j,j') \in \mathcal{J}}P\left[\left(c_{j}(\vec{x})=1\right) \cap \left(c_{j'}
(\vec{x})=1\right)\right]- \\
& \sum_{(j,j')\ne(k,k')\in\mathcal{J}}P\left[\left(c_{j}(\vec{x})=1\right) 
\cap \left(c_{j'}(\vec{x})=1\right)\right.\\
& \qquad \left. \cap \left(c_{k}(\vec{x})=1\right) \cap 
\left(c_{k'}(\vec{x})=1\right)\right] = 1,
\end{split}
\label{cond_3}
\end{equation}
\end{mycond}
where the overlap between two pairs of weak classifiers with labels $(j,j')$ and $(k,k')$ is given by the subtracted terms. Condition \ref{cond3} is illustrated in Fig.~\ref{cond_3_fig_2}.

It is possible to substitute a similar Condition \ref{cond3a} for the above Condition \ref{cond3} to create a different, yet also sufficient set of conditions for a completely accurate strong classifier. The number of weak classifiers 
required to satisfy the alternate set of conditions is expected {to} be smaller than the number 
required to satisfy the original three conditions. This is due to the fact that the 
modified conditions make use of one standalone weak classifier to cover a larger 
portion of the space correctly than is possible with a pair of weak classifiers.

\begin{figure}
\centering
\includegraphics[width=0.45\textwidth]{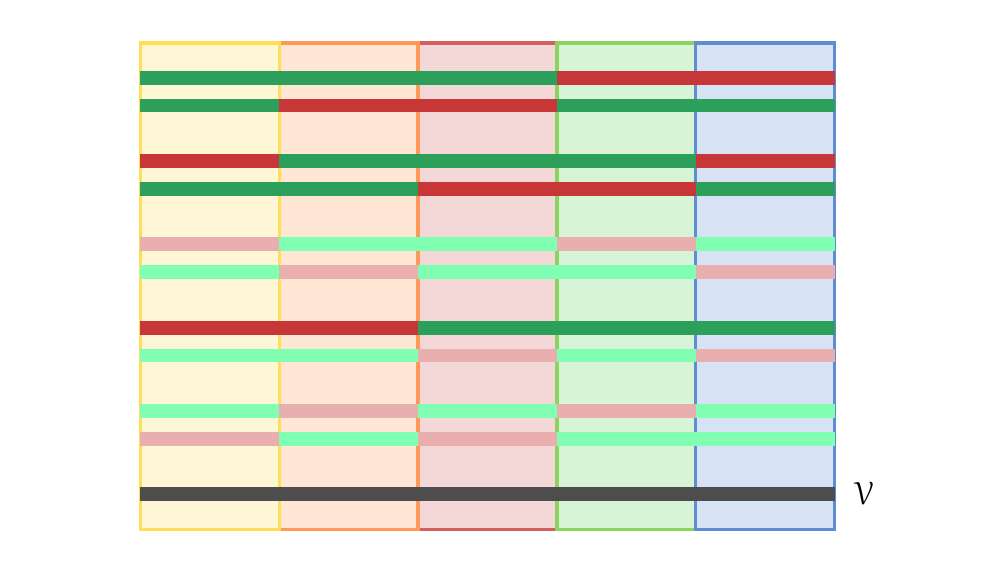}
\caption{Illustration of Condition~\ref{cond3a}. {Two} pairs and one single weak classifier form a completely accurate majority-vote strong classifier. The {two pairs cover 40\%} of the vector space with correct votes, and the single weak classifier (the first element of the {fourth} pair in Fig. \ref{cond_3_fig}; the faded-out classifiers in the {third, fourth, and fifth pairs are} omitted from this strong classifier) provides an extra correct vote to tip the balance in the remaining {60\%} to a correct overall classification.}
\label{cond_3a_fig}
\end{figure}

\addtocounter{myconda}{2}
\renewcommand{\themyconda}{\arabic{myconda}a}
\begin{myconda}
\label{cond3a}
\begin{equation}
\begin{split}
& \sum_{(j,j') \in J}P\left[\left(c_{j}(\vec{x})=1\right) \cap \left(c_{j'}
(\vec{x})=1\right)\right] + P\left[c_a(\vec{x})=1\right] - \\
& \sum_{(j,j')\ne(k,k')\in\mathcal{J}}
P\left[\left(c_{j}(\vec{x})=1\right)\cap \left(c_{j'}(\vec{x})=1\right) \right.\\
& \qquad \left. \cap \left(c_{k}
(\vec{x})=1\right) \cap \left(c_{k'}(\vec{x})=1\right)\right] - \\
& \sum_{(j,j')\in\mathcal{J}}P\left[\left(c_a(\vec{x})=1\right) \cap \left(c_{j}(\vec{x})=1
\right) \cap \left(c_{j'}(\vec{x})=1\right)\right] = 1
\end{split}
\label{cond_3a}
\end{equation}
\end{myconda}
This condition is illustrated in Fig.~\ref{cond_3a_fig}. Its interpretation is similar to that of Condition~\ref{cond3}, except that the standalone classifier with the subscript $a$ is added to the other classifier pairs, and its overlap with them is subtracted separately in the last line.

The perfect strong classifier can now be constructed from the weak classifiers in the set 
$\mathcal{J}$ defined by the conditions above. Define $\mathcal{J}_L$ as the set of all $j$ from pairs $(j,j')\in\mathcal{J}$. Similarly, define $\mathcal{J}_R$ as the set of all $j'$ from pairs $(j,j')\in\mathcal{J}$. Note that, since any pair for which $j=j'$ would not have minimum correctness overlap and therefore could not be in $\mathcal{J}$, it follow that $j\neq j'$ for all pairs $(j,j')$, i.e., $\mathcal{J}_L \cap \mathcal{J}_R = \emptyset$. The strong classifier is then (\ref{R}) with each $w_i$ being one of the elements of a pair, i.e.,

\begin{equation}
w_i=
\begin{cases}
1 & i\in(\mathcal{J}_L\cup\mathcal{J}_R)\\
0 & \text{otherwise}
\end{cases}
\label{cases}
\end{equation}

\subsection{Perfect strong classifier theorem}

We will now prove that any strong classifier satisfying Conditions \ref{cond1}-\ref{cond3}, or \ref{cond1}-\ref{cond3a}, is completely 
accurate.

\begin{mylemma}
\label{lemma1}
Assume Condition~\ref{cond1} and $(j,j')\in\J$. Then the sum of the correctness values of the corresponding weak classifiers is nonnegative everywhere with probability 1, namely
\begin{equation}
P\left[c_{j}(\vec{x})+c_{j'}(\vec{x}) \ge 0\right]=1
\end{equation}
for an input-output vector $\vec{x}\in \V$ selected uniformly at random.
\end{mylemma}
\begin{IEEEproof}
For any pair $(j,j')\in \mathcal{J}$ we have
\begin{equation}
\label{eqlemma1}
\begin{split}
&P\left[\left(c_{j}(\vec{x})=1\right){\cup}\left(c_{j'}(\vec{x})=1\right)\right]\\
& =P\left[c_{j}(\vec{x})=1\right]+P\left[c_{j'}(\vec{x})=1\right] \\
& \qquad -P\left[\left(c_{j}(\vec{x})=1\right){\cap}\left(c_{j'}(\vec{x})=1\right)\right]\\
& =1
\end{split}
\end{equation}
by Condition~\ref{cond2}. Eq.~\eqref{eqlemma1} means that at least one of the two weak classifiers evaluates to $1$. Since by definition $c_{i}(\vec{x}) \in \{-1,1\}$ $\forall i$, the sum is $2$ or $0$ with probability $1$, i.e.,
\ignore{
\begin{equation}
c_{i}(\vec{x}) \in \{-1,1\}
\quad\text{by definition}
\label{hvals}
\end{equation}
}
\begin{equation}
P\left[c_{j}(\vec{x})+c_{j'}(\vec{x})\in\{0,2\}\right]=1 .
\end{equation}
\end{IEEEproof}

Recall that if the majority of the votes given by the weak classifiers comprising a given strong classifier is true then the input-output vector being voted on receives a strong classification that is true [Eq.~\eqref{eq:perfect-weak}], and that if this is the case for all input-output vectors then the strong classifier is perfect [Eq.~\eqref{eq:perfect-strong}]. We are now in a position to state that this is the case with certainty provided the weak classifiers belong to the set $\J$ defined by the conditions given above.
\begin{mytheorem}
\label{theorem1}
A strong classifier comprised solely of a set of weak classifiers satisfying Conditions \ref{cond1}-\ref{cond3}
is perfect.
\end{mytheorem}
\begin{IEEEproof}
It suffices to show that the correctness sum is at least 2 
with probability 1 when Conditions \ref{cond1}-\ref{cond3}
are met, namely that
\begin{equation}
P\left[\sum_{(j,j')\in \mathcal{J}}\left(c_{j}(\vec{x})+c_{j'}(\vec{x})\right)
\ge 2\right]=1 .
\label{th1-eq}
\end{equation}
Now,
\bes
\bea
&&\!\!\!\!\!\!\! P\left[{\bigcup}_{(j,j')\in \mathcal{J}}\left(c_{j}(\vec{x})+c_{j'}(\vec{x})=2\right) \right] \notag \\
&& =P\left[{\bigcup_{(j,j')\in \mathcal{J}}}\left(c_{j}(\vec{x})=1\right)\cap\left(c_{j'}(\vec{x})=1\right)\right] \\
&&{\ge}\sum_{(j,j')\in \mathcal{J}}P\left[\left(c_{j}(\vec{x})=1\right)\cap\left(c_{j'}(\vec{x})=1\right)\right]\notag \\
&&-\sum_{(j,j')\ne (k,k')\in \mathcal{J}}P\left[\left(c_{j}(\vec{x})=1\right)
\cap\left(c_{j'}(\vec{x})=1\right) \right. \notag \\
\label{eq:46}
&&\qquad \left. \cap\left(c_{k}(\vec{x})=1\right) \cap\left(c_{k'}(\vec{x})=1\right)\right] \\
\label{th1-eq1}
&&=1\quad\text{by Cond.~\ref{cond3}.}
\eea
\ees
{where equality \eqref{th1-eq1} holds for the inequality \eqref{eq:46}\footnote{{This inequality reflects the fact that for $n$ overlapping sets, $P\left[\bigcup_{i=1}^{n}s_i\right]=\sum_{i=1}^{n}P[s_i]-\sum_{i\ne j}P[s_i\cap s_j] + \sum_{i\ne j \ne k}P[s_i\cap s_j \cap s_k] - \sum_{i\ne j\ne k \ne m}P[s_i\cap s_j \cap s_k \cap s_m]+\dots$ Each term is larger than the next in the series; $n+1$ sets cannot intersect where $n$ sets do not. Our truncation of the series is greater than or equal to the full value because we stop after a subtracted term.}} because the probability of an event cannot be greater than $1$.}

Thus, for any randomly selected vector $\vec{x}\in\mathcal{V}$, the correctness sum of at least one of the pairs is {$2$}, i.e.,
\begin{equation}
P\left[\exists (j,j')\in \mathcal{J}:\left(c_{j}(\vec{x})+c_{j'}(\vec{x})=2
\right)\right]=1 .
\label{th1-eq2}
\end{equation}
Lemma~\ref{lemma1} tells us that the correctness sum of each pair of weak classifiers is positive, while Eq.~\eqref{th1-eq2} states that for at least one pair this sum is not just positive but equal to $2$. Therefore the correctness sum of all weak classifiers in $\J$ is at least $2$, which is Eq.~\eqref{th1-eq2}.
\end{IEEEproof}

\begin{mytheorem}
\label{theorem3}
A strong classifier comprised solely of a set of weak classifiers satisfying Conditions \ref{cond1}, \ref{cond2}, and \ref{cond3a} is perfect.
\end{mytheorem}
\begin{IEEEproof}
It suffices to show that the correctness sum is 
at least 1
with probability 1 when Conditions \ref{cond1}, \ref{cond2}, and \ref{cond3a}
are met, namely that
\begin{equation}
P\left[\sum_{(j,j')\in \mathcal{J}}\left(c_{j}(\vec{x})+c_{j'}(\vec{x})\right)
+c_a(\vec{x})\ge 1\right]=1 .
\label{th3-eq2}
\end{equation}
We proceed similarly to the proof of Theorem~\ref{theorem1}.
\beq
\begin{split}
& P\left[\bigcup_{(j,j')\in \mathcal{J}}\left(c_{j}(\vec{x})+c_{j'}(\vec{x})=2\right)
\cup\left(c_{a}(\vec{x})=1\right)\right] \\
& =P\left[\bigcup_{(j,j')\in \mathcal{J}}\left(c_{j}(\vec{x})=1\right)\cap\left(c_{j'}
(\vec{x})=1\right)\cup\left(c_{a}(\vec{x})=1\right)\right]\\
& = \sum_{(j,j') \in J}P\left[\left(c_{j}(\vec{x})=1\right) \cap \left(c_{j'}
(\vec{x})=1\right)\right] + P\left[c_a(\vec{x})=1\right] \\
& - \sum_{(j,j')\ne(k,k')\in\mathcal{J}}
P\left[\left(c_{j}(\vec{x})=1\right)\cap \left(c_{j'}(\vec{x})=1\right) \right. \\
& \qquad\left. \cap \left(c_{k}
(\vec{x})=1\right) \cap \left(c_{k'}(\vec{x})=1\right)\right] \\
& - \sum_{(j,j')\in\mathcal{J}}P\left[\left(c_a(\vec{x})=1\right) \cap \left(c_{j}(\vec{x})=1
\right) \cap \left(c_{j'}(\vec{x})=1\right)\right]\\
&=1\quad\text{by Cond.~\ref{cond3a}}.
\end{split}
\eeq
Thus the correctness sum of at least one of the pairs together with the singled-out weak classifier is {greater than or equal to} $1$, i.e.,
\beq
P\left[\exists (j,j')\in \mathcal{J}:\left(c_{j}(\vec{x})+c_{j'}(\vec{x})=2
\right){\cup}\left(c_{a}(\vec{x})=1\right)\right]=1 .
\eeq
This result, together with Lemma~\ref{lemma1}, implies the correctness sum of all weak classifiers in $\J$ is at least $1$, which is Eq.~\eqref{th3-eq2}.
\end{IEEEproof}

\subsection{Imperfect strong classifier theorem}

Because the three conditions on the set $\J$ of weak classifiers guarantee a completely 
accurate strong classifier, errors in the strong classifier must mean that the 
conditions are violated in some way. For instance, Condition \ref{cond2} could be replaced by 
a weaker condition which allows for more than minimum overlap of vectors 
$\vec{x}$ categorized correctly by both weak classifiers in a pair.

\addtocounter{myconda}{-2}
\renewcommand{\themyconda}{\arabic{myconda}a}
\begin{myconda}
\label{cond2a}
If $(j,j') \in \mathcal{J}$ then
\beq
\begin{split}
& P\left[\left(c_{j}(\vec{x})=1\right)\cap \left(c_{j'}(\vec{x})=1
\right)\right] \\
& = P\left[c_{j}(\vec{x})=1\right] + P\left[c_{j'}(\vec{x})
=1\right] - 1 + \epsilon_{jj'} 
\label{cond_2a}
\end{split}
\eeq
for an input-output vector $\vec{x}\in \V$ selected uniformly at random.
\end{myconda}

The quantity $\epsilon_{jj'}$ is a measure of the ``overlap error". We can use it to prove relaxed versions of Lemma~\ref{lemma1} and Theorem~\ref{theorem1}.

\addtocounter{mylemma}{-1}
\renewcommand{\themylemma}{\arabic{mylemma}a}
\begin{mylemma}
\label{lemma1a}
Assume Condition~\ref{cond1} and $(j,j')\in\J$. Then the sum of the correctness values of the corresponding weak classifiers is nonnegative everywhere with probability $1-\epsilon_{jj'}$, namely
\begin{equation}
P\left[c_{j}(\vec{x})+c_{j'}(\vec{x})\ge 0 
\right]=1-\epsilon_{jj'} 
\end{equation}
for an input-output vector $\vec{x}\in \V$ selected uniformly at random.
\end{mylemma}
\begin{IEEEproof}
The proof closely mimics that of Lemma~\ref{lemma1}.
\beq
\begin{split}
&P\left[\left(c_{j}(\vec{x})=1\right){\cup}\left(c_{j'}(\vec{x})=1\right)\right]\\
& =P\left[c_{j}(\vec{x})=1\right]+P\left[c_{j'}(\vec{x})=1\right] \\
& \qquad -P\left[\left(c_{j}(\vec{x})=1\right){\cap}\left(c_{j'}(\vec{x})=1
\right)\right] \\
& =P\left[c_{j}(\vec{x})=1\right]+P\left[c_{j'}(\vec{x})=1\right] \\
& \qquad -P\left[c_{j}(\vec{x})=1\right]-P\left[c_{j'}(\vec{x})=1\right]
+1-\epsilon_{jj'}\\
&=1-\epsilon_{jj'}
\end{split}
\eeq
by Condition~\ref{cond2a}. As in the proof of Lemma~\ref{lemma1}, this implies
\beq
P\left[c_{j}(\vec{x})+c_{j'}(\vec{x})\in\{0,2\}\right]=1-\epsilon_{jj'}.
\eeq
\end{IEEEproof}
We can now replace Theorem~\ref{theorem1} by a lower bound on the success probability when Condition~\ref{cond2} is 
replaced by the weaker Condition~\ref{cond2a}. Let us first define an imperfect strong classifier as follows:
\begin{mydef}
A strong classifier is $\epsilon$-perfect if, for $\vec{x}\in\V$ chosen uniformly at random, it correctly classifies $\vec{x}$ [i.e., $Q_{\vec{w}}(\vec{x}) = y(\vec{x})$] with probability at least $1-\epsilon$.
\end{mydef}

\begin{mytheorem}
\label{theorem2}
A strong classifier comprised solely of a set of weak classifiers satisfying Conditions~\ref{cond1}, \ref{cond2a} and \ref{cond3} 
is $\epsilon$-perfect, where $\epsilon = \sum_{(j,j')\in \mathcal{J}}\epsilon_{jj'}$.
\end{mytheorem}
\begin{IEEEproof}
It suffices to show that the correctness sum is positive 
with probability 1 minus the sum of the overlap errors when Conditions~\ref{cond1}, \ref{cond2a} and \ref{cond3} 
are satisfied, namely
\begin{equation}
P\left[\sum_{(j,j')\in \mathcal{J}}c_{j}(\vec{x})+c_{j'}(\vec{x})
> 0\right]\ge 1-\sum_{(j,j')\in \mathcal{J}}\epsilon_{jj'}.
\label{1-eps}
\end{equation}
Now, by definition $c_{j}(\vec{x})+c_{j'}(\vec{x})\in\{-2,0,2\}$, and the correctness sum of at least one of the pairs must be negative in order for the correctness sum over all weak classifiers in $\J$ to be negative, so that
\bes
\bea
\label{prob0}
&&P\left[\sum_{(j,j')\in \mathcal{J}}c_{j}(\vec{x})+c_{j'}(\vec{x})
< 0\right]\\
&&\quad \le P\left[\exists (j,j')\in \mathcal{J}:c_{j}(\vec{x})+c_{j'}(\vec{x})
=-2\right] .
\label{prob1}
\eea
\ees
However, we also need to exclude the case of all weak classifier pairs summing to zero (otherwise the strong classifier can be inconclusive). This case is {partially} excluded by virtue of Condition~\ref{cond3}, which tells us that $\V$ as a whole is always covered by a correct majority vote. Formally, 
\beq
\begin{split}
& P\left[\sum_{(j,j')\in \mathcal{J}}c_{j}(\vec{x})+c_{j'}(\vec{x})= 0\right] \\
&= P\left[\bigcap_{(j,j')\in \mathcal{J}}\left(c_{j}(\vec{x})+c_{j'}(\vec{x})\right)= 0\right]\\
&=1-P\left[\exists (j,j')\in \mathcal{J}:c_{j}(\vec{x})+c_{j'}(\vec{x})>0\right]\\
&=0,
\end{split}
\eeq
where in the last equality we invoked the calculation leading from Eq.~\eqref{th1-eq1} to Eq.~\eqref{th1-eq2}, which only required Condition~\ref{cond3}. Alternatively, we could use Condition~\ref{cond3a} to prove that $P\left[\sum_{(j,j')\in \mathcal{J}}c_{j}(\vec{x})+c_{j'}(\vec{x}){+c_a(\vec{x})}= 0\right]=0$. {There is another way for the classifier to return an inconclusive result: if one weak classifier pair has a correctness sum of $2$ and another weak classifier pair has a correctness sum of $-2$. This case is included in the bound in Eq. \ref{prob1} because one of the weak classifier pairs in this scenario has a negative correctness sum.} We can thus conclude that the strict inequality in Eq.~\eqref{prob0} can be replaced by $\leq$.

Now, the probability of there being one weak classifier pair such as in Eq.~\eqref{prob1} cannot be greater than the probability of at least one of the pairs having a negative correctness sum, which in turn---by the union bound---cannot be greater than the sum of such probabilities:
\bea
\text{Eq.~}\eqref{prob1} &\leq& P\left[\bigcup_{(j,j')\in \mathcal{J}}\left(c_{j}(\vec{x})+c_{j'}(\vec{x})=-2\right)\right] \notag\\
&\le&\sum_{(j,j')\in \mathcal{J}}P\left[c_{j}(\vec{x})+c_{j'}(\vec{x})=-2\right]\\
&=&\sum_{(j,j')\in \mathcal{J}}\epsilon_{jj'} \notag ,
\eea
where the last equality follows from Lemma~\ref{lemma1a}. This proves Eq.~\eqref{1-eps}.
\end{IEEEproof}

It is interesting to note that---as alluded to in this proof---if we were to drop Conditions~\ref{cond3} and \ref{cond3a}, then Eq.~\eqref{1-eps} would become $P\left[\sum_{(j,j')\in \mathcal{J}}c_{j}(\vec{x})+c_{j'}(\vec{x})
\geq 0\right]{\ge 1-}\sum_{(j,j')\in \mathcal{J}}\epsilon_{jj'}$ (note the change from $>$ to $\geq$), so that Theorem~\ref{theorem2} would change to a statement about inconclusive $\epsilon$-perfect strong classifiers, which can---with finite probability---yield a ``don't-know" answer. This may be a useful tradeoff if it turns out to be difficult to construct a set of weak classifiers satisfying Condition \ref{cond3} or \ref{cond3a}.

\subsection{An alternate weight optimization problem}
\label{sec:alternate}

The conditions and results {established} in the previous subsection for correctness of the strong classifier suggest the creation of an alternate weight optimization problem to select the weak classifiers that will be included in the final majority vote, replacing the optimization problem of {Section~\ref{sec:relaxed}}. The new optimization problem is defined over the space of \emph{pairs} of weak classifiers, rather than singles, which can be constructed using elements of the set $\AtA$, with $\A$ as defined in Condition~\ref{cond1}. We define the \emph{ideal} pair weight as

\begin{equation}
\tilde{w}_{ij}=
\begin{cases}
1 & (i,j)\in{\mathcal{J}\times\mathcal{J}}\\
0 & \text{otherwise}
\end{cases},
\label{eq:wdef}
\end{equation}
Since we do not know the set $\J$ \textit{a priori}, we shall define a QUBO whose solutions $w_{ij}\in\{0,1\}$, with $(i,j)\in \AtA$, will be an approximation to the ideal pair weights $\tilde{w}_{ij}$. In the process, we shall map the pair weight bits $w_{ij}$ to qubits. Each $w_{ij}$ determines whether its corresponding pair of weak classifiers, $h_i$ and $h_j$, will be included in the new strong classifier, which can thus be written as:

\begin{align}
Q^{\rm pair}(\vec{x}) & =\mathrm{sign}\left[  R_{\vec{w}^{\rm pair}}(\vec{x})\right]\nonumber\\
& =\mathrm{sign}\left[  \sum_{(i,j)\in\AtA}w_{ij}\left(h_{i}(\vec{x})+h_j(\vec{x})\right)\right]
\label{Qpair}
\end{align}

Recall that we do not know the $w_{ij}$ {\it a priori}; they are found in our approach via the solution of a QUBO, which we set up as follows:
\begin{align}
& \vec{w}_{\rm pair}^{\rm opt}=\arg\min_{\vec{w}}\left[\sum_{(i,j)\in{\AtA}}
\alpha_{ij}w_{ij} \right.\nonumber\\
& \qquad\left. +\sum_{(i,j)\ne(k,l)\in
{\AtA}}J_{ijkl}w_{ij}w_{kl}\right],
\end{align}
where the second term is a double sum over all sets of unequal pairs. The solution of this QUBO will provide us with an approximation to the set $\J$, which yields the desired set of weak classifiers as in Eq.~\eqref{cases}. 
Sparsity can be enforced as in Eq.~\eqref{QUBO} by replacing $\alpha_{ij}$ with $\alpha_{ij}{+}\lambda$, where $\lambda>0$, i.e., by including a penalty proportional to $\| \vec{w}\| _{0}$.

The terms $\alpha_{ij}$ and $J_{ijkl}$ reward 
compliance with Conditions \ref{cond2} and \ref{cond3}, respectively.
To define $\alpha_{ij}$, we 
first define the {modified correctness} function
$c'_i:\T\mapsto \{0,1\}$, where $\T$ is the training set \eqref{trainset}:

\begin{equation}
c'_i(\vec{x}_{s},y_{s})=\frac{1}{2}\left(h_i(\vec{x}_s)y_s+1\right)=\begin{cases}
1 & h_i(\vec{x}_s)=y_s\\
0 & h_i(\vec{x}_s)\neq y_s
\end{cases}
\end{equation}
Below we write $c'_i(s)$ in place of $c'_i(\vec{x}_{s},y_{s})$ for notational simplicity.
The term $\alpha_{ij}$ rewards the pairing of weak classifiers which classify the 
minimal number of vectors $\vec{x}$ incorrectly at the same time, as specified by 
Condition~\ref{cond2}. Each pair included gains negative weight for the training set vectors its 
members classify correctly, but is also given a positive penalty for any vectors 
classified incorrectly by both weak classifiers at once:

\begin{equation}
\alpha_{ij}=-\frac{1}{S}\sum_{s=1}^S\left[c'_i(s)+c'_j(s)-
\left(1-c'_i(s)\right)\left(1-c'_j(s)\right)\right]
\end{equation}

The term $J_{ijkl}$ penalizes the inclusion of pairs that are too similar to each 
other, as codified in Condition~\ref{cond3}. This is accomplished by assigning a
positive weight for each vector that is classified correctly by two pairs at once:

\begin{equation}
J_{ijkl}=\frac{1}{S}\sum_{s=1}^Sc'_i(s)c'_j(\vec{x_s})c'_k(s)c'_l(s)
\end{equation}

We now have a QUBO for the alternate weight optimization problem. This can be 
translated to the Ising Hamiltonian as with the original optimization problem in 
Section \ref{QUBO2Ising}. We again map from our QUBO variables $w_{ij}$ to 
variables $q_{ij}=2(w_{ij}-1/2)$, yielding the following optimization function:

\beq
\begin{split}
\vec{q}_{\rm pair}^{\rm opt} 
& =\arg\min_{\vec{q}}\left[\frac{1}{2}\sum_{(i,j)\in
{\AtA}} \beta_{ij} q_{ij} \right.\\
& \qquad\left. +
\frac{1}{4}\sum_{(i,j)\ne(k,l)\in
{\AtA}}J_{ijkl}q_{ij}q_{kl}\right],
\end{split}
\eeq
where
\beq
\beta_{ij} = \alpha_{ij}+\frac{1}{2}\left(\sum_{(k,l)\in
{\AtA};(k,l)\neq (i,j)}J_{ijkl}+J_{klij}\right).
\eeq
Constant terms were omitted because they have 
no bearing on the minimization. This optimization function is now suitable for direct 
translation to the final Hamiltonian for an AQC:

\beq
\begin{split}
H_F & =\frac{1}{2}\sum_{(i,j)\in
{\AtA}} \beta_{ij}
Z_{ij} \\
& \qquad +
\frac{1}{4}\sum_{(i,j)\ne(k,l)\in
{\AtA}}J_{ijkl}Z_{ij}Z_{kl}.
\end{split}
\eeq

The 
qubits now represent weights on pairs rather than on an individual classifier. $Z_{ij}$ is therefore the Pauli $\sigma_z$ operator on the qubit assigned to the pair $(i,j) \in \AtA$.
Using $|\A|^2$ qubits, this approach will give the optimal combination of weak 
classifier pairs over the training set according to the conditions set forth previously.

\section{Using strong classifiers in quantum-parallel}
\label{sec_use}%

Now let us suppose that we have already trained our strong classifier and
found the optimal weight vector $\vec{w}^{\rm opt}$ or $\vec{w}^{\rm opt}_{\rm pair}$. For simplicity we shall henceforth limit our discussion to $\vec{w}^{\rm opt}$. We can use 
the trained classifier to
classify new input-output pairs $\vec{x}\notin \mathcal{T}$ 
to decide whether they are correct
or erroneous. In this section we shall address the question of
how we can further obtain a
quantum speedup in exhaustively testing \emph{all} exponentially many
($2^{N_{\rm in}+N_{\rm out}}$) input-output pairs $\vec{x}$. The key observation in
this regard is that if we can formulate software error testing as a minimization problem
over the space $\mathcal{V}$ of all input-output pairs $\vec{x}$, then an AQC\ algorithm
will indeed perform a quantum-parallel search over this entire space,
returning as the ground state an erroneous state.

\subsection{Using two strong binary classifiers to detect errors}

Recall that we are concerned with the detection of vectors $\vec{x}\in \V$ that are erroneous and implemented [Eq.~\eqref{eq:impl-err}]. 
To accomplish this, we use two strong classifiers. The \emph{specification classifier} 
is the binary classifier developed in Section \ref{sec_train}. {Ideally,} it behaves as follows:
\begin{equation}
Q_{\vec{w}}(\vec{x})=\begin{cases}
1 & \vec{x}\in{\hat{S}}\\
-1 & \vec{x}\notin{\hat{S}}
\end{cases}
\label{q_w}
\end{equation}
The second classifier, which we will call the \emph{implementation classifier}, determines 
whether or not an input-output vector is in the program as implemented. It is constructed 
in the same way as $Q_{\vec{w}}(\vec{x})$, but with its own appropriate training set. Ideally, it behaves as follows:
\begin{equation}
T_{\vec{z}}(\vec{x})=\begin{cases}
1 & \vec{x}\notin{S}\\
-1 & \vec{x}\in{S}
\end{cases}
\label{t_z}
\end{equation}
The four possible combinations represented by Eqs.~\eqref{q_w} and \eqref{t_z} correspond to the four cases covered by Definitions\ref{def:impl-err}-\ref{def:unimpl-err}.
The worrisome input-output vectors, those that are erroneous and implemented, cause both classifiers to evaluate to $-1$.

\subsection{Formal criterion}

As a first step, suppose we use the optimal weights vector in the original
strong specification classifier. We then have, from ~(\ref{Q}),
\begin{equation}
Q^{\rm opt}(\vec{x})=\mathrm{sign}\left[  R_{\vec{w}^{\rm opt}}(\vec{x})\right]
=\mathrm{sign}\left[  \sum_{i=1}^{N}w_{i}^{\rm opt}h_{i}(\vec{x})\right]
\label{Qopt}%
\end{equation}
This, of course, is imprecise since our adiabatic algorithm
solves a relaxed optimization problem (i.e., returns $\vec{w}^{\rm opt}$, not
$\vec{w}^{\prime {\rm opt}}\,$), but we shall assume that the replacement is sufficiently 
close to the true optimum for our purposes. 
With this caveat, Eq.~\eqref{Qopt} is the optimal strong specification classifier for a given input-output vector $\vec{x}$,
with the classification of $\vec{x}$ as erroneous if $Q^{\rm opt}(\vec{x})=-1$ or as
correct if $Q^{\rm opt}(\vec{x})=+1$.

The strong implementation classifier is constructed similarly to the specification 
classifier:
\begin{equation}
T^{\rm opt}(\vec{x})=\mathrm{sign}\left[U_{\vec{z}^{\rm opt}}(\vec{x})\right]
=\mathrm{sign}\left[\sum_{i=1}^{N}z_{i}^{\rm opt}h_{i}(\vec{x})\right]
\label{Topt}%
\end{equation}
Here, $h_{i}$ are the same weak classifiers as those used to train the specification
classifier, but $T^{\rm opt}$ is constructed independently from a 
training set $\T^\prime$ which may or may not overlap with $\T$.
This training set is labeled according to the possibility or impossibility of producing the input-output pairs in $\T^\prime$ from the implemented program. The result of this optimization is the weight vector $\vec{z}^{\rm opt}$.

Given the results of the classifiers $Q^{\rm opt}(\vec{x})$ and $T^{\rm opt}(\vec{x})$ for any 
vector $\vec{x}$, the V\&V task of identifying whether or not $\vec{x}\in(S\cap\neg\hat{S})$ 
reduces to the following. Any vector $\vec{x}$ is flagged as erroneous and implemented 
if $Q^{\rm opt}(\vec{x})+T^{\rm opt}(\vec{x})=-2$. 
We stress once more that, due to our use of the relaxed optimization to solve for $\vec{w}^{\rm opt}$ and $\vec{z}^{\rm opt}$,  a flagged $\vec{x}$ may in fact be neither erroneous nor implemented, i.e., our procedure is susceptible to both false positives and false negatives.

\subsection{Relaxed criterion}
\label{sec:relaxed_testing}

{As was the case with} Eq.~\eqref{w-opt}, $Q^{\rm opt}+T^{\rm opt}$ is {unfortunately} not directly implementable in AQC, but a simple
relaxation is. The trick is again to remove the sign function, this time from
~(\ref{Qopt}) and (\ref{Topt}), and consider the sum of the two classifiers'
majority vote functions directly as an energy function:
\begin{equation}
C^{\rm opt}(\vec{x})=R_{\vec{w}^{\rm opt}}(\vec{x})+U_{\vec{z}^{\rm opt}}(\vec{x})
\label{Copt}
\end{equation}
The combination of the two classifiers gives different results for vectors falling 
under each of the Definitions from Section \ref{sets_section}.

\emph{Case 1:} $\vec{x}\notin\hat{S}$ and $\vec{x}\in S$\\
The vector $\vec{x}$ is an error implemented in the program and manifests a software error. These vectors gain negative 
weight from both classifiers $R_{\vec{w}^{\rm opt}}$ and $U_{\vec{z}^{\rm opt}}$. Vectors 
falling under this definition should receive the lowest values of $C^{\rm opt}$,
if any such vectors exist.

\emph{Case 2:} $\vec{x}\in\hat{S}$ and $\vec{x}\in S$\\
The vector $\vec{x}$ satisfies the don't-worry condition, that is, it is a correct input-output string, 
part of the ideal program 
$\hat{P}$. In this case, $R_{\vec{w}^{\rm opt}}>0$ and $U_{\vec{z}^{\rm opt}}<0$. In the programs
quantum V\&V is likely to be used for, with very infrequent, elusive errors, the 
specification and implementation will be similar and the negative weight of 
$U_{\vec{z}^{\rm opt}}<0$ should be moderated enough by the positive influence of 
$R_{\vec{w}^{\rm opt}}>0$ that don't-worry vectors should not populate the lowest-lying states.

\emph{Case 3:} $\vec{x}\in\hat{S}$ and $\vec{x}\notin S$\\
The input portion of the vector $\vec{x}$ is a don't-care condition. It does not 
violate any program specifications, but is not important enough to be specifically 
addressed in the implementation. This vector will gain positive weight from both 
$R_{\vec{w}^{\rm opt}}$ and $U_{\vec{z}^{\rm opt}}$ and should therefore never be misidentified 
as an error.

\emph{Case 4:} $\vec{x}\notin\hat{S}$ and $\vec{x}\notin S$\\
The vectors $\vec{x}$ in this category would be seen as erroneous by the program 
specification - if they ever occurred. Because they fall outside the program 
implementation $S$, they are not the errors we are trying to find. This case is 
similar to the don't-worry situation in that the two strong classifiers will have opposite signs, 
in this case $R_{\vec{w}^{\rm opt}}<0$ and $U_{\vec{z}^{\rm opt}}>0$. By the same argument as 
Definition 2, Definition 4 vectors should not receive more negative values of 
$C^{\rm opt}$ than the targeted errors.

Having examined the values of $C^{\rm opt}(\vec{x})$ for the relevant categories of $\vec{x}$, 
we can formulate error detection as the following minimization problem:%
\begin{equation}
\vec{x}_{e}=\arg\min_{\vec{x}}C^{\rm opt}(\vec{x}).
\label{x_b}%
\end{equation}
Suppose the algorithm returns a solution $\vec{x}_{e}$ ($e$ for ``error''). We
then need to test that it is indeed an error, which amounts to checking that it
behaves incorrectly when considered as an input-output pair in the
program implementation $P$. Note that testing that $R_{\vec{w}^{\rm opt}}(\vec
{x}_{e})<0$ is insufficient, since our procedure involved a sequence of relaxations.

\subsection{Adiabatic implementation of the relaxed criterion}
\label{impl}%

In order to implement the error identification strategy (\ref{x_b}) we need to consider
\beq
C^{\rm opt}(\vec{x})=\sum_{i=1}^{N}(w_{i}^{\rm opt}+z_{i}^{\rm opt})h_{i}(\vec{x})
\label{eq:Copt}
\eeq 
as an
energy function. We then consider $C^{\rm opt}(\vec
{x})$ as the final Hamiltonian $H_F$ for an AQC, with Hilbert space spanned by the
basis $\{|\vec{x}\rangle\}$. The AQC will then find the state which
minimizes $C^{\rm opt}(\vec{x})$ out of all $2^{N_{\rm in}+N_{\rm out}}$ basis
states and thus identify an error candidate. Because the AQC always returns \emph{some} error candidate, our procedure never generates false negatives. However, Cases 2 and 4 would correspond to false positives, if an input-output vector satisfying either one of these cases is found as the AQC output.

We can rely on the fact that the AQC\ actually returns a (close
approximation to the) Boltzmann distribution
\begin{equation}
\Pr[\vec{x}]=\frac{1}{Z}\exp[-C^{\rm opt}(\vec{x})/(k_{B}T)],
\end{equation}
where $k_{B}$ is the Boltzmann constant, $T$ is the temperature, and
$Z=\sum_{\vec{x}}\exp[-C^{\rm opt}(\vec{x})/(k_{B}T)]$ is the partition
function. For sufficiently low temperature this probability distribution is
sharply peaked around the ground state, with contributions from the first few
excited states. Thus we can expect that even if there is a low-lying state that
has been pushed there by only one of the two binary classifiers $Q^{\rm opt}$ or $T^{\rm opt}$, the AQC will return a nearby state
which is both erroneous and implemented some of the time and  
an error will still be 
detected. Even if the undesirable state [$\vec{x}\in\hat{S}$ and $\vec{x}\in S$, or $\vec{x}\notin\hat{S}$ and $\vec{x}\notin S$] is the ground state, and hence 
all erroneous states [$\vec{x}\notin\hat{S}$ and $\vec{x}\in S$] are excited
states, their lowest energy member will be found with a probability that is
$e^{-\Delta(t_F)/(k_{B}T)}$ smaller than the unlooked-for state, where
$\Delta(t_F)$ is the energy gap to the first excited state at the end of the computation. Provided $k_{B}T$ and $\Delta(t_F)$ are of the same order, this probability will be appreciable.

To ensure that errors which are members of the training set are never identified as ground states we construct the training set $\T$ so that it only includes correct states, i.e., $y_s=+1$ $\forall s$. This has the potential drawback that the classifier never trains directly on errors. It is in principle possible to include errors in the training set ($y_s =-1$) by adding another penalty term to the strong classifier which directly penalizes such training set members, but whether this can be done without introducing many-body interactions in $H_F$ is a problem that is beyond the scope of this work.

\subsection{Choosing the weak classifiers}
\label{weaks}%

Written in the form $\sum_{i=1}^{N}(w_{i}^{\rm opt}+z_{i}^{\rm opt})h_{i}(\vec{x})$, the energy
function $C^{\rm opt}(\vec{x})$ is too general, since we haven't yet 
specified the weak classifiers\ $h_{i}(\vec{x})$. However, we are free to
choose these so as to mold $C^{\rm opt}(\vec{x})$ into a Hamiltonian
that is physically implementable in AQC. 

\begin{table*}[htbp]
\centering
\begin{tabular}{|l|l|p{2.5in}|l|}
\hline
Function \# & Boolean Logic & Intermediate Form & Implementation Form \\ \hline
$i=0$ & $x_{i_{3}}==0$ & not applicable & $-Z_{i_3}$ \\ \hline
$i=1$ & $x_{i_3}==\overline{\left( x_{i_1}\vee x_{i_2}\right) }$ & $4\left(x_{i_1}x_{i_2}x_{i_3}-x_{i_1}x_{i_3}-x_{i_2}x_{i_3}\right)-2\left(x_{i_1}x_{i_2}-x_{i_1}-x_{i_2}-x_{i_3}\right)-1$ & $Z_a\otimes Z_{i_3} -Z_{i_1}\otimes Z_{i_3} -Z_{i_2}\otimes Z_{i_3} $ \\ \hline
$i=2$ & $x_{i_3}==\overline{x_{i_1}}\wedge x_{i_2}$ & $4(-x_{i_1}x_{i_2}x_{i_3}+x_{i_2}x_{i_3})+2(-x_{i_3}+x_{i_1}x_{i_2}-x_{i_2})+1$ & $-Z_a\otimes Z_{i_3}+Z_{i_2}\otimes Z_{i_3} - Z_{i_3}$ \\ \hline
$i=3$ & $x_{i_3}==\overline{x_{i_1}}$ & not applicable & $-Z_{i_3}\otimes Z_{i_1}$ \\ \hline
$i=4$ & $x_{i_3}==x_{i_1}\wedge \overline{x_{i_2}}$ & $4(x_{i_1}x_{i_3}-x_{i_1}x_{i_2}x_{i_3})-2(x_{i_1}-x_{i_1}x_{i_2}+x_{i_3})+1$ & $Z_{i_1}\otimes Z_{i_3}-Z_{a}\otimes Z_{i_3}-Z_{i_3}$ \\ \hline
$i=5$ & $x_{i_3}==\overline{x_{i_2}}$ & not applicable & $-Z_{i_3}\otimes Z_{i_2}$ \\ \hline
$i=6$ & $x_{i_3}==x_{i_1}\oplus x_{i_2}$ & $-8x_{i_1}x_{i_2}x_{i_3} + 4(x_{i_1}x_{i_3}+x_{i_2}x_{i_3}+x_{i_1}x_{i_2})-2(x_{i_1}+x_{i_2}+x_{i_3})+1$ & $-2Z_a \otimes Z_{i_3}+Z_{i_1}\otimes Z_{i_3}+Z_{i_2}\otimes Z_{i_3}-Z_{i_3}$ \\ \hline
$i=7$ & $x_{i_3}==\overline{\left( x_{i_1}\wedge x_{i_2}\right) }$ & $-4x_{i_1}x_{i_2}x_{i_3}+2(x_{i_3}+x_{i_1}x_{i_2})-1$ & $-Z_a\otimes Z_{i_3}$ \\ \hline
$i=8$ & $x_{i_3}==x_{i_1}\wedge x_{i_2}$ & $4x_{i_1}x_{i_2}x_{i_3}-2(x_{i_3}+x_{i_1}x_{i_2})+1$ & $Z_a\otimes Z_{i_3}$ \\ \hline
$i=9$ & $x_{i_3}==\overline{\left( x_{i_1}\oplus x_{i_2}\right) }$ & $8x_{i_1}x_{i_2}x_{i_3} -4(x_{i_1}x_{i_3}+x_{i_2}x_{i_3}+x_{i_1}x_{i_2})+2(x_{i_1}+x_{i_2}+x_{i_3})-1$ & $2Z_a \otimes Z_{i_3}-Z_{i_1}\otimes Z_{i_3}-Z_{i_2}\otimes Z_{i_3}+Z_{i_3}$ \\ \hline
$i=10$ & $x_{i_3}==x_{i_2}$ & not applicable & $Z_{i_3}\otimes Z_{i_2}$ \\ \hline
$i=11$ & $x_{i_3}==\overline{x_{i_1}}\vee x_{i_2}$ & $-4(x_{i_1}x_{i_3}-x_{i_1}x_{i_2}x_{i_3})+2(x_{i_1}-x_{i_1}x_{i_2}+x_{i_3})-1$ & $-Z_{i_1}\otimes Z_{i_3}+Z_{a}\otimes Z_{i_3}+Z_{i_3}$ \\ \hline
$i=12$ & $x_{i_3}==x_{i_1}$ & not applicable & $Z_{i_3}\otimes Z_{i_1}$ \\ \hline
$i=13$ & $x_{i_3}==x_{i_1}\vee \overline{x_{i_2}}$ & $-4(-x_{i_1}x_{i_2}x_{i_3}+x_{i_2}x_{i_3})-2(-x_{i_3}+x_{i_1}x_{i_2}-x_{i_2})-1$ & $Z_a\otimes Z_{i_3}-Z_{i_2}\otimes Z_{i_3} + Z_{i_3}$ \\ \hline
$i=14$ & $x_{i_3}==x_{i_1}\vee x_{i_2}$ & $-4\left(x_{i_1}x_{i_2}x_{i_3}-x_{i_1}x_{i_3}-x_{i_2}x_{i_3}\right)+2\left(x_{i_1}x_{i_2}-x_{i_1}-x_{i_2}-x_{i_3}\right)+1$ & $-Z_a\otimes Z_{i_3} +Z_{i_1}\otimes Z_{i_3} +Z_{i_2}\otimes Z_{i_3} $ \\ \hline
$i=15$ & $x_{i_3}==1$ & not applicable & $Z_{i_3}$ \\ \hline
\end{tabular}
\caption{All $16$ Boolean functions $f_i$ of two binary variables, and their implementation form in terms of the Pauli matrices $Z_{i_j}$ acting on single qubits or pairs of qubits $j\in\{1,2,3\}$. The subscript $a$ in the Implementation Form column denotes an ancilla qubit, tied to qubits $i_1$ and $i_2$ via $x_a=x_{i_1}x_{i_2}$, used to reduce all qubit interactions to at most two-body.}
\label{dictionary_table}
\end{table*}%

Suppose, e.g., that $h_{i}%
(\vec{x})$ measures a Boolean relationship defined by a function $f_{i}:\{0,1\}^{\ell}\mapsto \{0,1\}$ 
between several bits of the input-output vector; 
$x_{k}=\mathrm{bit}_{k}(\vec{x})$, the $k$th bit of $\vec{x}\in \V$. For example,
\begin{equation}
h_{i}(\vec{x})=(x_{i_{3}}==f_{i}(x_{i_{1}},x_{i_{2}})),
\label{wc1}%
\end{equation}
where ``$a==b$'' evaluates to $1$ if $a=b$ or to $0$ if $a\neq b$. Here $i_{1}$ and $i_{2}$
are the positions of two bits from the input vector $\vec
{x}_{\rm in}$ and $i_{3}$ 
is the position of a bit from the output vector $\vec{x}_{\rm out}$, 
so that $h_{i}$
measures a correlation between inputs and outputs.
The choice of this particular form for the weak classifiers is physically motivated, as it corresponds to at most three-body interactions between qubits, which can all be reduced to two-body interaction by the addition of ancilla qubits (see below). Let us enumerate these weak classifiers.
The number of different
Boolean functions $f_i$ is $2^{2^{\ell}}$
\cite{Slepian:53}.\footnote{Any Boolean function of $\ell$ variables can be uniquely expanded in the form $f_i(x_1,\dots,x_\ell) = \sum_{\alpha=0}^{2^\ell-1} \epsilon_{i\alpha} s_{\alpha}$, where $\epsilon_{i\alpha}\in\{0,1\}$ and $s_\alpha$ are the $2^\ell$ ``simple" Boolean functions $s_0 = x_1 x_2 \cdots x_\ell$, $s_1 = x_1 x_2 \cdots \overline{x_\ell}$, $\dots$, $s_{2^\ell-1} = \overline{x_1}\, \overline{x_2} \cdots \overline{x_\ell}$, where $\overline{x}$ denotes the negation of the bit $x$. Since each $\epsilon_{i\alpha}$ can assume one of two values, there are $2^{2^\ell}$ different Boolean functions.} Much more efficient representations are possible under reasonable assumptions \cite{Bryant:86}, but for the time being we shall not concern ourselves with these.
In the example of the classifier~\eqref{wc1}
there are $N_{\rm in}(N_{\rm in}-1)$ input bit combinations
for each of the $N_{\rm out}$ output bits.
The number of different Boolean functions in this example, where $\ell=2$, is $2^{2^{2}}=16$.
Thus the dimension of the ``dictionary'' of weak classifiers is
\beq
N=16 N_{\rm in}(N_{\rm in}-1)N_{\rm out}
\eeq
for the case of Eq.~\eqref{wc1}.

{We wish to find a two-local quantum implementation for each $h_i(\vec{x})$ in the dictionary. {It is possible to find a two-local implementation for any three-local Hamiltonian using so-called ``perturbation gadgets'', or three ancilla bits for each three-local term included \cite{PhysRevA.77.062329}, but rather than using the general method we rely on a special case which will allow us to use only one ancilla bit per three-local term.} We first devise an \textit{intermediate form} function using products of the same bits $x_i\in\{0,1\}$ used to define the logical behavior of each weak classifier. This function will have a value of $1$ when the Boolean relationship specified for $h_i(\vec{x})$ is true, and $-1$ otherwise. For example, consider function number 8, $x_{i_3}==x_{i_1}\wedge x_{i_2}$, the AND function. Its intermediate form is $4x_{i_1}x_{i_2}x_{i_3}-2\left(x_{i_3}+x_{i_1}x_{i_2}\right)+1$. For the bit values $(x_{i_1},x_{i_2},x_{i_3})=(0,0,0)$, the value of the intermediate function is $1$, and the Boolean form is true: $0$ AND $0$ yields $0$. If instead we had the bit values $(x_{i_1},x_{i_2},x_{i_3})=(0,0,1)$, the intermediate form would yield $-1$, and the Boolean form would be false, because the value for $x_{i_3}$ does not follow from the values for $x_{i_1}$ and $x_{i_2}$.}

The two-body implementation form is obtained in two steps from the intermediate form. First, an ancilla bit tied to the product of the two input bits, $x_a=x_{i_1}x_{i_2}$, is substituted into any intermediate form expressions involving three-bit products. This is permissible because such an ancilla can indeed be created by introducing a penalty into the final Hamiltonian for any states in which the ancilla bit is not equal to the product $x_{i_1}x_{i_2}$. We detail this method below. Then, the modified intermediate expression is translated into a form that uses bits valued as $x'_i\in\{-1,1\}$ rather than $x_i\in\{0,1\}$ using the equivalence $x_i=2x'_i-1$. The modified intermediate form is now amenable to using the implemented qubits. Note that the Pauli matrix $Z_{i}$ acts on a basis ket $|\vec{x}\rangle$ as 
\beq
Z_{i}|\vec{x}\rangle=(-1)^{\mathrm{bit}_{i}(\vec{x})}|\vec{x}\rangle.
\eeq
This means that we can substitute $Z_{i}$ for $x'_i$ and $Z_{i}\otimes Z_{j}$ for $x'_ix'_j$ in the intermediate form, resulting in the implementation form given in Column 4 of Table \ref{dictionary_table}.  Some weak classifiers do not involve three-bit interactions. Their implementation forms were devised directly, a simple process when there is no need for inclusion of an ancilla.

{We have reduced the dictionary functions from three-bit to two-bit interactions by adding an ancilla bit to represent the product of the two input bits involved in the function. Therefore, the maximum number of qubits needed to implement this set of weak classifiers on a quantum processor 
is $Q=N_{\rm in}+N_{\rm out}+N_{\rm in}^2$.} In practice, it is likely to be significantly 
less because not every three-bit correlation will be relevant to a given 
classification problem. 

Let us now discuss how the penalty function is {introduced.} For example, consider {again} the implementation of weak classifier function $i=8$, whose intermediate form involves three-qubit products, which we reduced to two-qubit interactions by including $x_a$.

We {ensure that $x_a$ does indeed represent the product it is intended to} by making the function a sum of two terms: the product of the ancilla qubit and the remaining qubit from the original product, and a term that adds a penalty if the ancilla is not in fact equal to the product of the two qubits it is meant to represent, in this case {$f_{\text{penalty}}=x_{i_1}x_{i_2}-2(x_{i_1}+x_{i_2})x_a+3x_a$. In the case where $(x_{i_1},x_{i_2},x_a)=(1,0,0)$, $f_{\text{penalty}}=0$, but in a case where $x_a$ does not represent the intended product {such as} $(x_{i_1},x_{i_2},x_a)=(1,0,1)$, $f_{\text{penalty}}=1$. In fact, the penalty function behaves as follows:}
\begin{equation}
f_{\text{penalty}}=\begin{cases}
0 & x_a=x_{i_1}x_{i_2} \\
\text{positive} & \text{otherwise}
\end{cases}
\end{equation}
In the end, we have the modified intermediate form $f_8 = 4x_{a}x_{i_3}-2\left(x_{i_3}+x_{a}\right)+1+f_{\text{penalty}}$, which involves only two-qubit interactions. This would be implemented on the quantum computer as the sum of two Hamiltonian terms:
\begin{equation}
H_8 = Z_a\otimes Z_{i_3},
\end{equation}
from the implementation column of Table \ref{dictionary_table}, and
\begin{align}
H_{\text{penalty}}(i_1,i_2)= & \frac{1}{4}Z_{i_1}\otimes Z_{i_2}-\frac{1}{2}Z_{i_1}\otimes Z_a-\frac{1}{2}Z_{i_2}\otimes Z_a \nonumber\\
& - \frac{1}{4}Z_{i_1}-\frac{1}{4}Z_{i_2}+\frac{1}{2}Z_a+\frac{3}{4},
\end{align}
the implementation form of $f_{\text{penalty}}$, so a Hamiltonian to find input-output vectors classified negatively by this weak classifier would be
\begin{equation}
H_{\text{weak}} = H_8 + H_{\text{penalty}}(i_1,i_2).
\end{equation}

{When the strong classifier is implemented as a whole, multiple weak classifiers with weight $1$ may use the same two input bits, and therefore share an ancilla bit that is the product of those input bits. When this is the case, it is sufficient to add the penalty function to the final Hamiltonian once, though the ancilla is used multiple times.}

{The inclusion of ancilla qubits tied to products of other qubits and their associated penalties need not interfere with the solution of the 
{V\&V} problem, although the ancilla penalty terms must appear in the same final Hamiltonian as this optimization. If the ancilla penalty terms are made reasonably large, they will put any states in which the ancillas do not represent their intended products (states which are in fact outside of $\V$) far above the levels at which errors are found. For instance, consider an efficient, nearly optimal strong classifier closely approximating the conditions set forth in Section \ref{conditions}. Such a classifier makes its decision on the strength of two simultaneously true votes. If two such classifiers are added together, as in the verification problem, the lowest energy levels will have an energy near $-4$. If the penalty on a forbidden ancilla state is more than a reasonable $4$ units, such a state should be well clear of the region where errors are found.}

This varied yet correlation-limited set of weak classifiers 
fits nicely with the idea of tracking
intermediate spaces [Eq.~(\ref{int})], where we can use an intermediate space
$\mathcal{I}_{j}$ to construct a set of weak classifiers feeding into the next
intermediate space $\mathcal{I}_{j+1}$. This is further related to an obvious
objection to the above classifiers, which is that they ignore
any correlations involving four or more bits, without one-, two-, or three-bit
correlations. By building a hierarchy of weak classifiers, for intermediate
spaces, such correlations can hopefully be accounted for as they build
up by keeping track instead of
one-, two-, and three-bit terms as the program runs.

\subsection{QUBO-AQC quantum parallel testing}
 
With the choice of Boolean functions for the weak classifiers, the quantum
implementation of the energy function $C^{\rm opt}(\vec{x})$ [Eq.~\eqref{eq:Copt}]
becomes
\begin{equation}
H_{F}^{\mathrm{test}}=\sum_{i=1}^{N}(w_{i}^{\rm opt}+z_{i}^{\rm opt}){H_i + \sum_{j\ne k}H_{\text{penalty}}(j,k)} ,
\label{H_F}
\end{equation}
where {$H_i$} denotes the implemented form given in the third column of Table~\ref{dictionary_table}, and the indices $j,k\in\{1,\dots,N_{\text{in}}\}$ denote all possible pairings of input qubits tied to ancillas. The ground state of $H_{F}^{\mathrm{test}}$, which corresponds to the optimal
weight sets $w_{i}^{\rm opt}$ and $z_{i}^{\rm opt}$ derived from the set of weak classifiers 
detailed in Subsection \ref{weaks}, is an erroneous state, which, by construction, is not a member of the training set $\T$.

How do we construct the AQC\ such that all input-output pairs $\vec{x}$ are
tested in parallel? This is a consequence of the adiabatic interpolation Hamiltonian \eqref{interp}, and in particular the initial Hamiltonian $H_I$ of the type given in Eq.~\eqref{H_I}. The ground state of this positive semi-definite $H_I$ is an equal superposition over all input-output vectors, i.e., $H_I\sum_{\vec{x}\in\V}|\vec{x}\rangle = 0$, and hence
when we implement the AQC every possible $\vec{x}$ starts out as
a candidate for the ground
state. The final (Boltzmann)\ distribution of observed states strongly favors
the manifold of low energy states, and by design these will be implemented erroneous states,
it they exist.

\section{Sample problem implementation}
\label{simulation_section}

In order to explore the practicality of our two-step adiabatic quantum approach to 
finding software errors, we have applied the algorithm to a program of limited size
containing a logical error. We did this by calculating the results of the algorithm 
assuming perfect adiabatic quantum optimization steps on a processor with few 
$(N<30)$ available qubits. Preliminary characterizations of the accuracy achievable 
using such an algorithm given a set of weak classifiers with certain characteristics 
are also presented.

\subsection{The Triplex Monitor Miscompare problem}
\label{triplex}%

The problem we chose to implement is a toy model of 
program design practices used in mission critical software systems.\footnote{We are grateful to Greg Tallant from the Lockheed Martin Corporation for providing us with this problem as an example of interest in flight control systems.}
This program monitors a set of three redundant variables $\{A_t,B_t,C_t\}$ for internal consistency. The 
variables could represent, e.g., sensor inputs, control signals, or particularly important internal program values.
If one value is different from the other 
two over a predetermined number of snapshots in time $t$, a problem 
in the system is indicated and the value of the two consistent redundant variables 
is propagated as correct. Thus the program is supposed to implement a simple majority-vote error-detection code.

We consider only the simplest case of two time snapshots, i.e., $t=1,2$. 
As just explained, a correct implementation of the monitoring routine should fail a 
redundant variable $A$, $B$, or $C$ if that \emph{same} variable miscompares with both of the other variables in each of the two time frames. The erroneous implemented program  
we shall consider has the 
logical error that, due to a mishandled internal implementation of the miscompare 
tracking over multiple time frames, it fails a redundant variable any time there has 
been a miscompare in both time frames, even if the miscompare implicated a \emph{different} variable in each time frame. 

In order to facilitate quantum V\&V using the smallest 
possible number of qubits, we assume the use of classical preprocessing to reduce the 
program to its essential structure. The quantum algorithm does not look 
at the values of the three redundant variables in each time frame. Instead, it 
sees three logical bits per snapshot, telling it whether each pair of variables is equal. This 
strategy is also reflected in the program outputs, which are three logical bits 
indicating whether or not each redundant variable is deemed correct by the 
monitoring routine. Thus there are nine logical bits, as specified in Table~\ref{tab3}.

\begin{table}[htbp]
\centering
\begin{tabular}{| l | l |}
\hline
Bit & Significance \\ \hline
$x_1$ & $A_1 \ne B_1$ \\ \hline
$x_2$ & $B_1 \ne C_1$ \\ \hline
$x_3$ & $A_1 \ne C_1$ \\ \hline
$x_4$ & $A_2 \ne B_2$ \\ \hline
$x_5$ & $B_2 \ne C_2$ \\ \hline
$x_6$ & $A_2 \ne C_2$ \\ \hline
$x_7$ & $A$ failed \\ \hline
$x_8$ & $B$ failed \\ \hline
$x_9$ & $C$ failed \\ \hline
\end{tabular}
\caption{Logical bits and their significance in terms of variable comparison in the Triplex Miscompare problem.}
\label{tab3}
\end{table}

In terms of Boolean logic, the two behaviors are as 
follows:

\underline{Program Specification}
\begin{subequations}
\begin{equation}
x_7 = x_1 \land x_3 \land x_4 \land x_6,
\end{equation}
\begin{equation}
x_8 = x_1 \land x_2 \land x_4 \land x_5,
\end{equation}
\begin{equation}
x_9 = x_2 \land x_3 \land x_5 \land x_6,
\end{equation}
\end{subequations}
i.e., a variable is flagged as incorrect if and only if it has miscompared with all other 
variables in all time frames.

\underline{Erroneous Program Implementation}
\begin{subequations}
\begin{equation}
x_7 = ((x_1 \land x_2) \lor (x_2 \land x_3) \lor (x_1 \land x_3)) \land x_4 \land x_6,
\end{equation}
\begin{equation}
x_8 = ((x_1 \land x_2) \lor (x_2 \land x_3) \lor (x_1 \land x_3)) \land x_4 \land x_5,
\end{equation}
\begin{equation}
x_9 = ((x_1 \land x_2) \lor (x_2 \land x_3) \lor (x_1 \land x_3)) \land x_5 \land x_6,
\end{equation}
\end{subequations}
i.e., a variable is flagged as incorrect if it miscompares 
with the other variables in the final time frame and if any variable has miscompared 
with the others in the previous time frame.

\subsection{Implemented algorithm}
\label{algorithm}

\begin{figure*}
\centering
\subfloat{\includegraphics[width=0.45\textwidth]{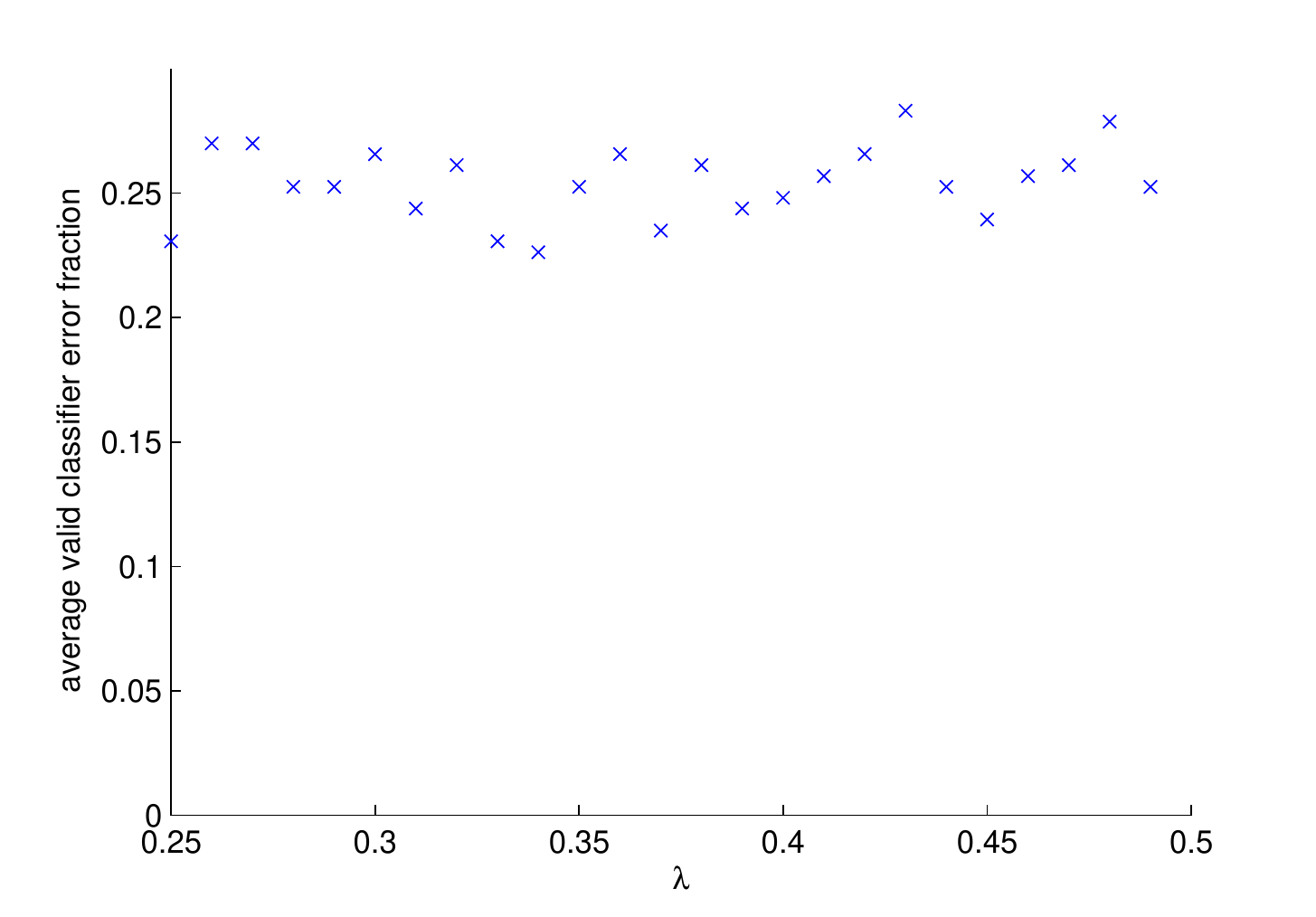}}
\subfloat{\includegraphics[width=0.45\textwidth]{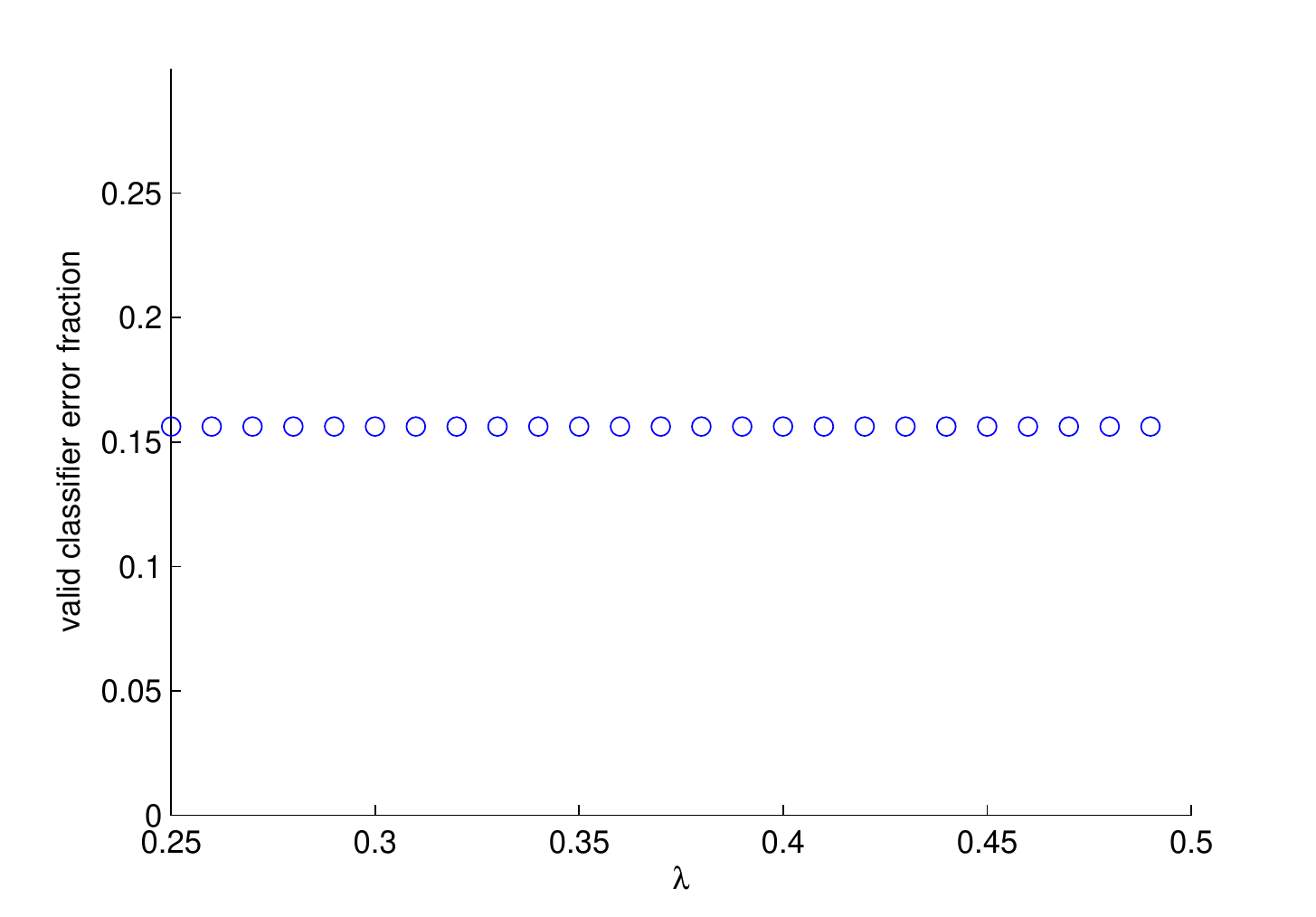}}
\caption{Error fractions in 16-member specification classifier calculations; Left: average over 50. Right: best of 50.}
\label{50_correct}
\end{figure*}

\begin{figure*}
\centering
\subfloat{\includegraphics[width=0.31\textwidth, angle=90]{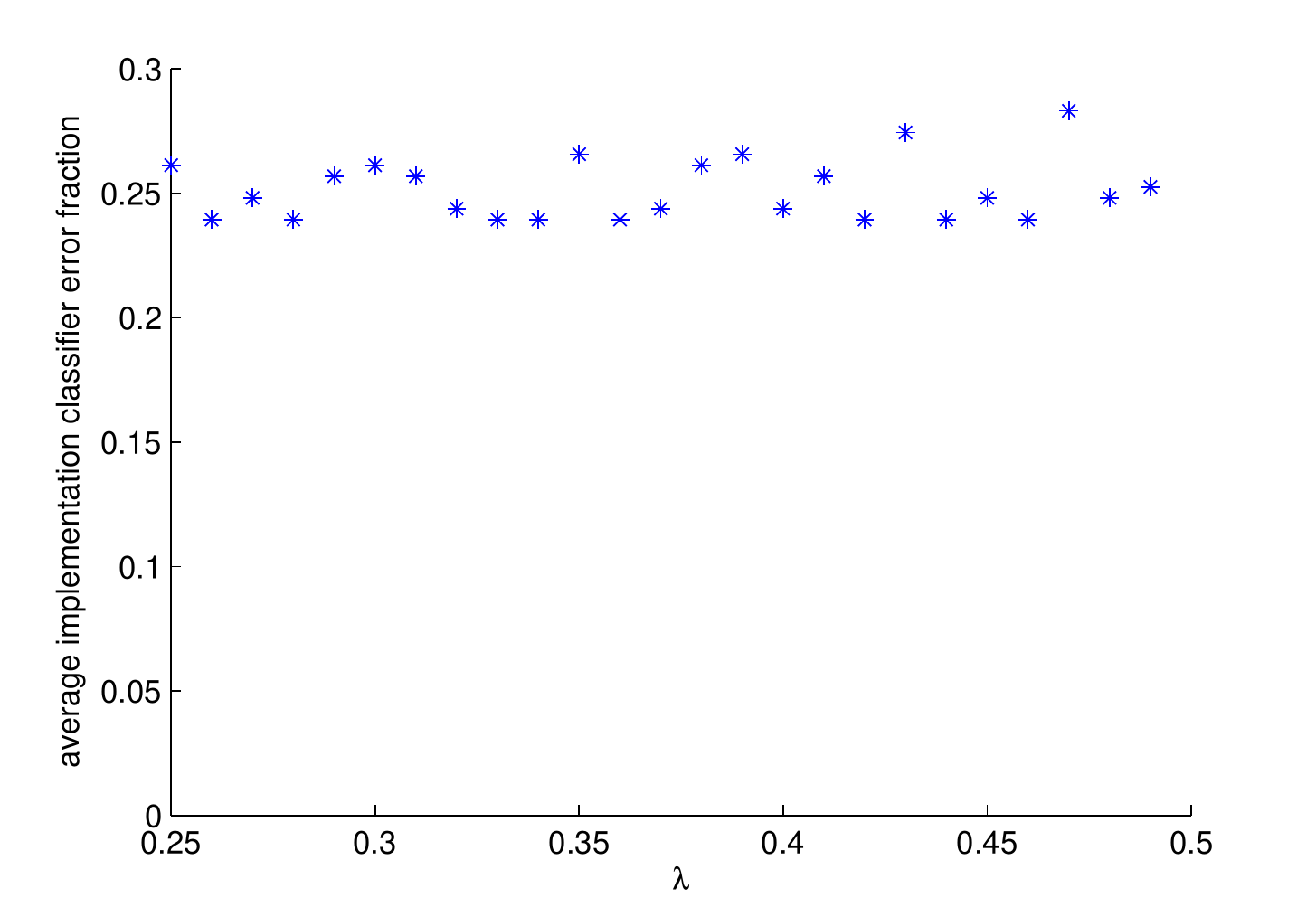}}
\subfloat{\includegraphics[width=0.45\textwidth]{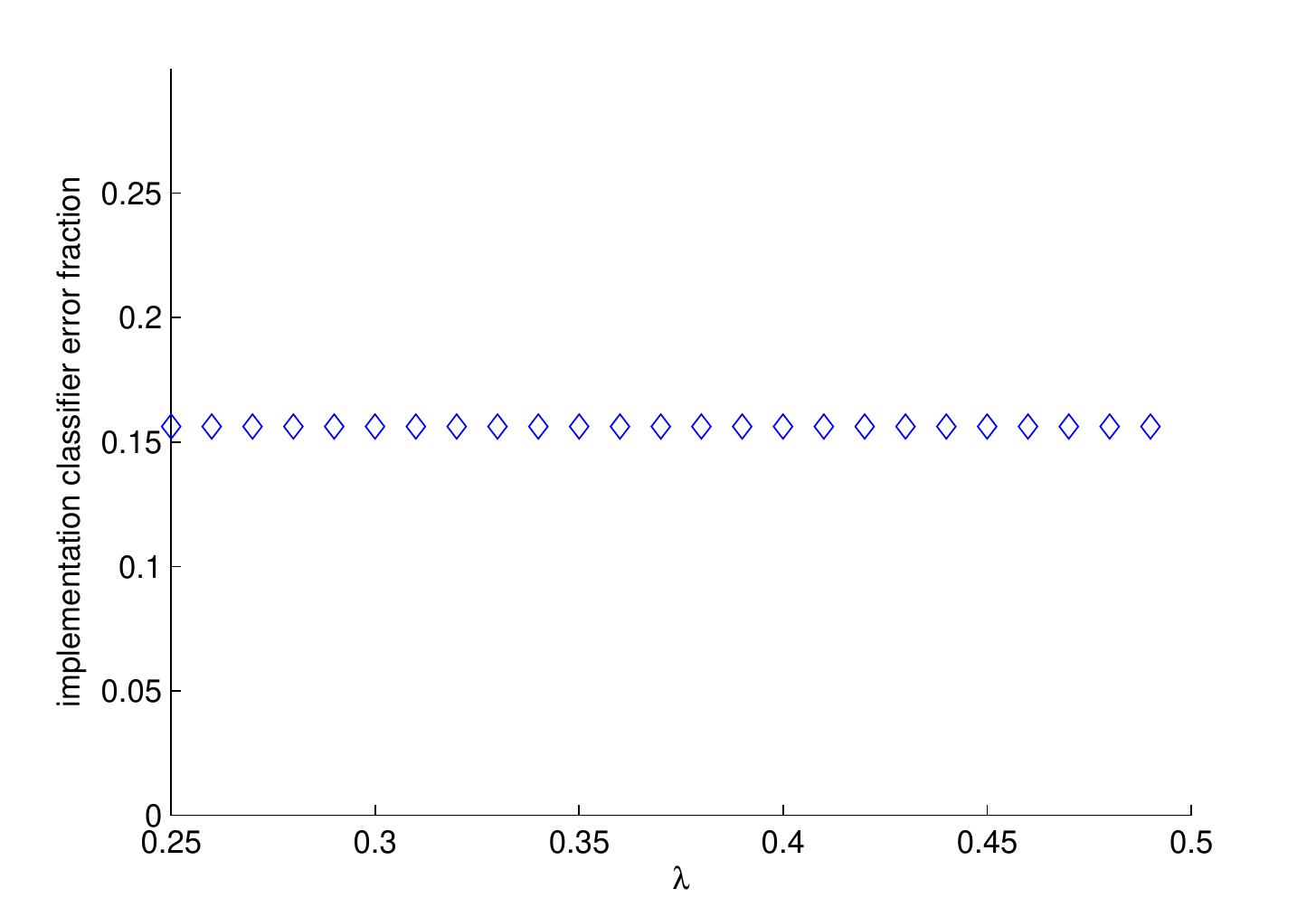}}
\caption{Error fractions in 16-member implementation classifier calculations; Left: average over 50. Right: best of 50.}
\label{50_impl}
\end{figure*}

The challenges before us are to train classifiers to 
recognize the behavior of both the program specification and the erroneous 
implementation, and then to use those classifiers to find the errors. These 
objectives have been programmed into a hybrid quantum-classical algorithm using the
quantum techniques described in Sections \ref{sec_train} and \ref{sec_use} and 
classical strategy refinements based on characteristics of available resources 
(for example, the accuracy of the set of available weak classifiers). The performance 
of this algorithm has been tested through computational studies using a classical 
optimization routine in place of adiabatic quantum optimization calls.

The algorithm takes as its inputs two training sets, one for the specification 
classifier and one for the implementation classifier. The two strong classifiers 
are constructed using the same method, one after the other, consulting the 
appropriate training set.

When constructing a strong classifier, the algorithm first evaluates the performance 
of each weak classifier in the dictionary over the training set. Weak classifiers 
with poor performance, typically those with over 40\% error, are discarded. The 
resulting, more accurate dictionary is fed piecewise into the quantum optimization 
algorithm.

Ideally, the adiabatic quantum optimization using the final Hamiltonian \eqref{Hf}
would take 
place over the set of all weak classifiers in the modified, more accurate dictionary. 
However, the reality of quantum computation for some time to come is that the 
number of qubits available for processing will be smaller than the number of 
weak classifiers in the accurate dictionary. This problem is addressed by selecting 
random groups of $Q$ classifiers (the number of available qubits) to be optimized 
together. An initial random group of $Q$ classifiers is selected, the optimal 
weight vector $\vec{q}^{\rm opt}$ is calculated {by classically finding the ground state of $H_F$}, and the weak classifiers which receive 
weight $0$ are discarded. The resulting spaces are filled in with weak classifiers 
randomly selected from the set of those which have not yet been considered, until 
all $Q$ classifiers included in the optimization return a weight of $1$. This procedure 
is repeated until all weak classifiers in the accurate dictionary have been 
considered, at which time the most accurate group of $Q$ generated in this manner is 
accepted as the strong classifier for the training set in question. Clearly, alternative strategies for combining subsets of $Q$ weak classifiers could be considered, such as genetic algorithms, but this was not attempted here.

Both the specification and implementation strong classifiers are generated in this 
way, resulting in
\begin{equation}
R_{\vec{w}^Q}(\vec{x})=\sum_{i=1}^Nw_i^Qh_i(\vec{x})
\label{R_Q}
\end{equation}
\begin{equation}
T_{\vec{z}^Q}(\vec{x})=\sum_{i=1}^Nz_i^Qh_i(\vec{x})
\label{T_Q}
\end{equation}
where $w_i^Q$ and $z_i^Q$ take the value 1 if the corresponding weak classifier 
$h_i(\vec{x})$ is selected using the iterative procedure described in the 
preceding paragraph, and are zero otherwise. This is the same structure as that 
seen in Eqs.~(\ref{Qopt}) and (\ref{Topt}), but with different vectors $\vec{w}$ and $\vec{z}$ 
due to the lack of available qubits to perform a global optimization over the 
accurate dictionary.

The two strong classifiers of Eqs.~(\ref{R_Q}) and (\ref{T_Q}) are summed as 
in Eq.~(\ref{Copt}) to create a final energy function that will push errors to 
the bottom part of the spectrum. This is translated to a final Hamiltonian $H_F$ as in 
Eq.~({\ref{H_F}) and the result of the optimization (i.e., the ground state of this $H_F$) is returned as the 
error candidate. This portion of the algorithm makes it crucial to employ intelligent 
classical preprocessing in order to keep the length of the input and output vectors as small as 
possible, because each bit in the input-output vector corresponds to a qubit%
, and the classical cost of finding the ground state of $H_F$ grows exponentially with the number of qubits.

\subsection{Simulation results}

\begin{figure*}
\centering
\subfloat{\includegraphics[width=0.45\textwidth]{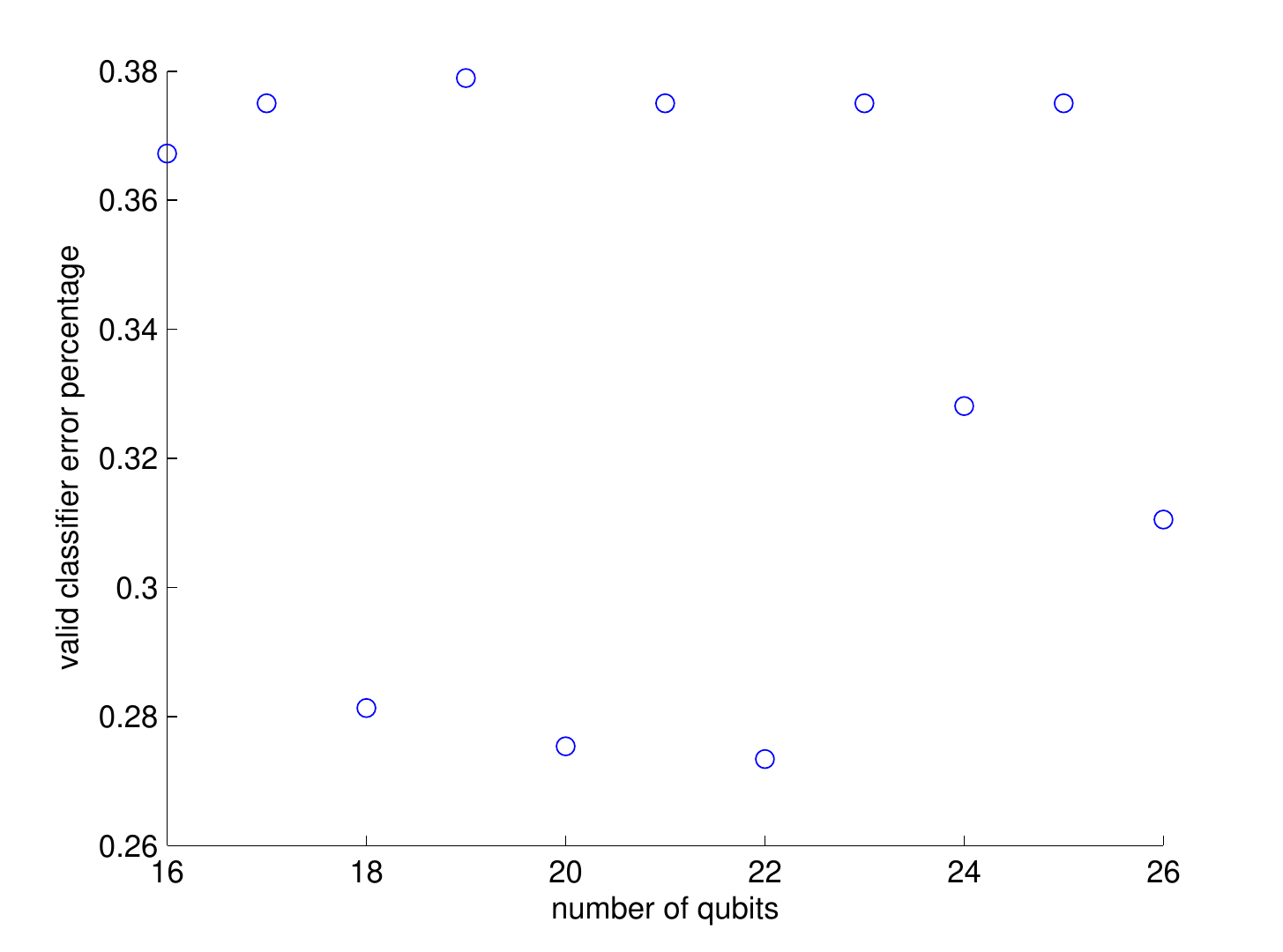}}
\subfloat{\includegraphics[width=0.45\textwidth]{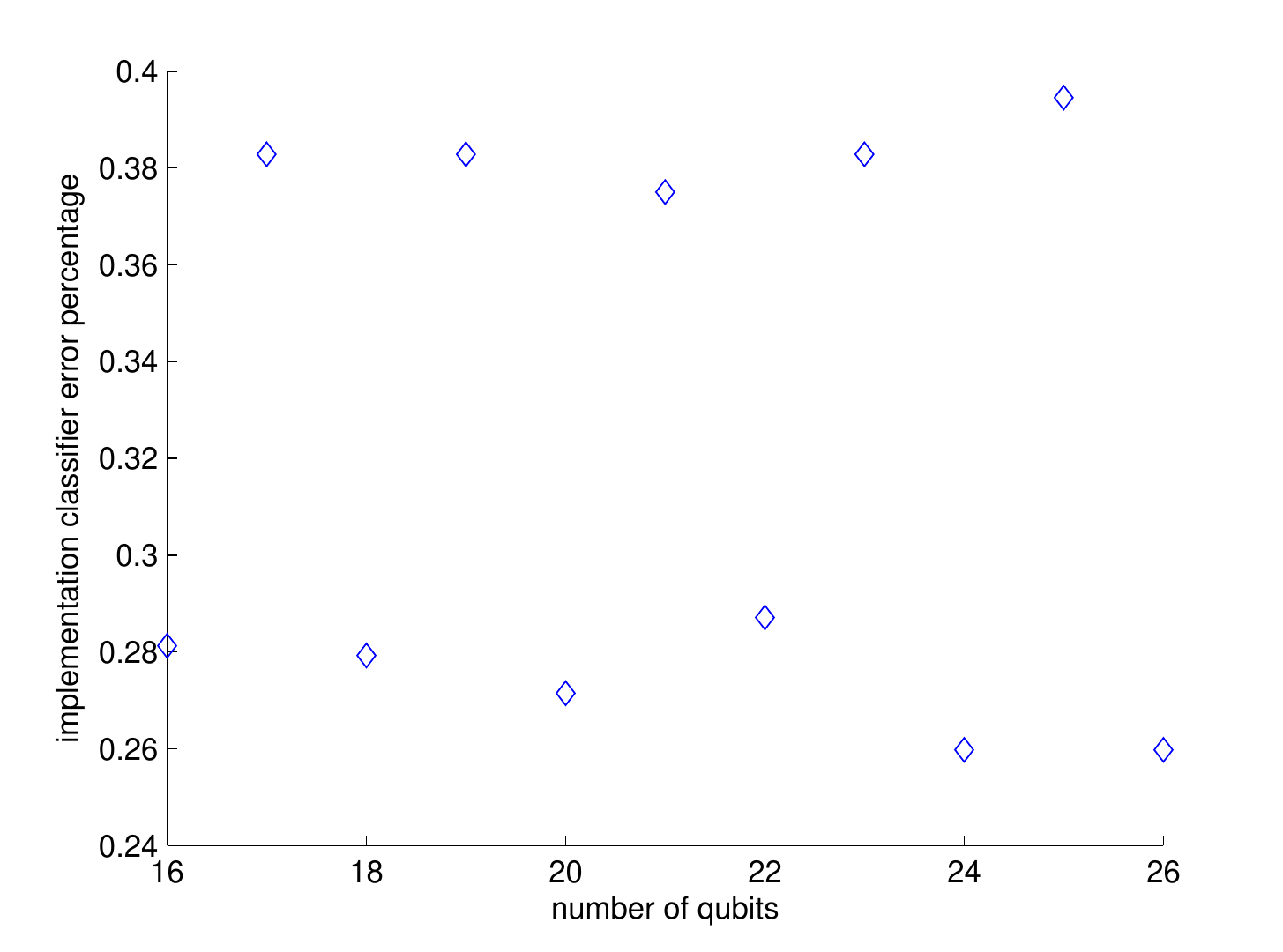}}
\caption{Error fractions of specification (left) and implementation (right) classifiers,
for an increasing number of qubits.}
\label{qubits_graph}
\end{figure*}

Our simulation efforts have focused on achieving better accuracy from the two 
strong classifiers. If the strong classifiers are not highly accurate, the 
second part of the algorithm, the quantum-parallel use of the classifiers, 
will not produce useful results.

In the interest of pushing the limits of accuracy of the strong classifiers, 
some simulations were performed on the miscompare problem in a single 
time frame. Under this simplification, the program specification and 
implementation are identical (the error arises over multiple time frames), and 
indeed the numerical results will show that the results for the two 
classifiers are the same (see Figs.~\ref{50_correct} and \ref{50_impl}, right).

The algorithm described in Subsection \ref{algorithm} was run $50$ times, each 
time producing two strong classifiers comprising $16$ or fewer weak classifier 
members. The figure of $16$ qubits was chosen because it allowed the computations 
to be performed in a reasonable amount of time on a desktop computer while still allowing for some 
complexity in the makeup of the strong classifiers. This set of $50$ complete 
algorithmic iterations was performed for $26$ values of $\lambda$, the sparsity 
parameter introduced in Eq.~(\ref{w-opt}). The average percentage of error 
for both strong classifiers was examined, as was the best error fraction 
achieved in the $50$ iterations. These two quantities are defined as follows:

\bea
\label{avg_err}
{\rm err}_{\rm avg} &=& \frac{1}{50}\sum_{i=1}^{50}L_i(\vec{w}^{\rm opt}) \\
\label{min_err}
{\rm err_{\min} } &=& 
\min_i L_i(\vec{w}^{\rm opt}),
\eea
where $L$ is the function that counts the total number of incorrect classifications, Eq.~\eqref{L}.
The weight vector $\vec{z}^{\rm opt}$ can be substituted for $\vec{w}^{\rm opt}$ in Eqs. \eqref{avg_err} and \eqref{min_err} if the strong classifier being analyzed is the implementation rather than the specification classifier.

Both the average and minimum error for the specification and implementation classifiers are plotted in Figs.~\ref{50_correct} and \ref{50_impl}, respectively,
as a function of $\lambda$.

As shown in Figs.~\ref{50_correct} and \ref{50_impl}, while the average percent 
error for both classifiers 
hovered around 25\%, the best percent error was consistently just below 16\% 
for both the specification and implementation classifiers. The consistency suggests 
two things: that the randomness of the algorithm can be tamed by looking for 
the best outcome over a limited number of iterations, and that the sparsity 
parameter, $\lambda$, did not have much effect on classifier accuracy.

Noting in particular the lack of dependency on $\lambda$, we move forward to 
examine the results of simulations on more difficult and computationally 
intensive applications of the algorithm. These results address the triplex 
monitor miscompare problem exactly as described in subsection \ref{triplex} 
and increase the number of qubits as far as $26$. The error fractions of the 
best strong classifiers found, defined as

\begin{equation}
\textrm{err}_{\min}(Q)= \min_i L_i(\vec{w}^{\rm opt}) \qquad i\in\{1,\dots,n_{\rm sim}(Q)\}
\end{equation}
where $n_{\rm sim}(Q)$ is the number of simulations performed at $Q$ qubits, are plotted in Fig. \ref{qubits_graph} as a function of the number of qubits
allowed in the simulation.

\begin{figure*}
\centering
\includegraphics[width=\textwidth]{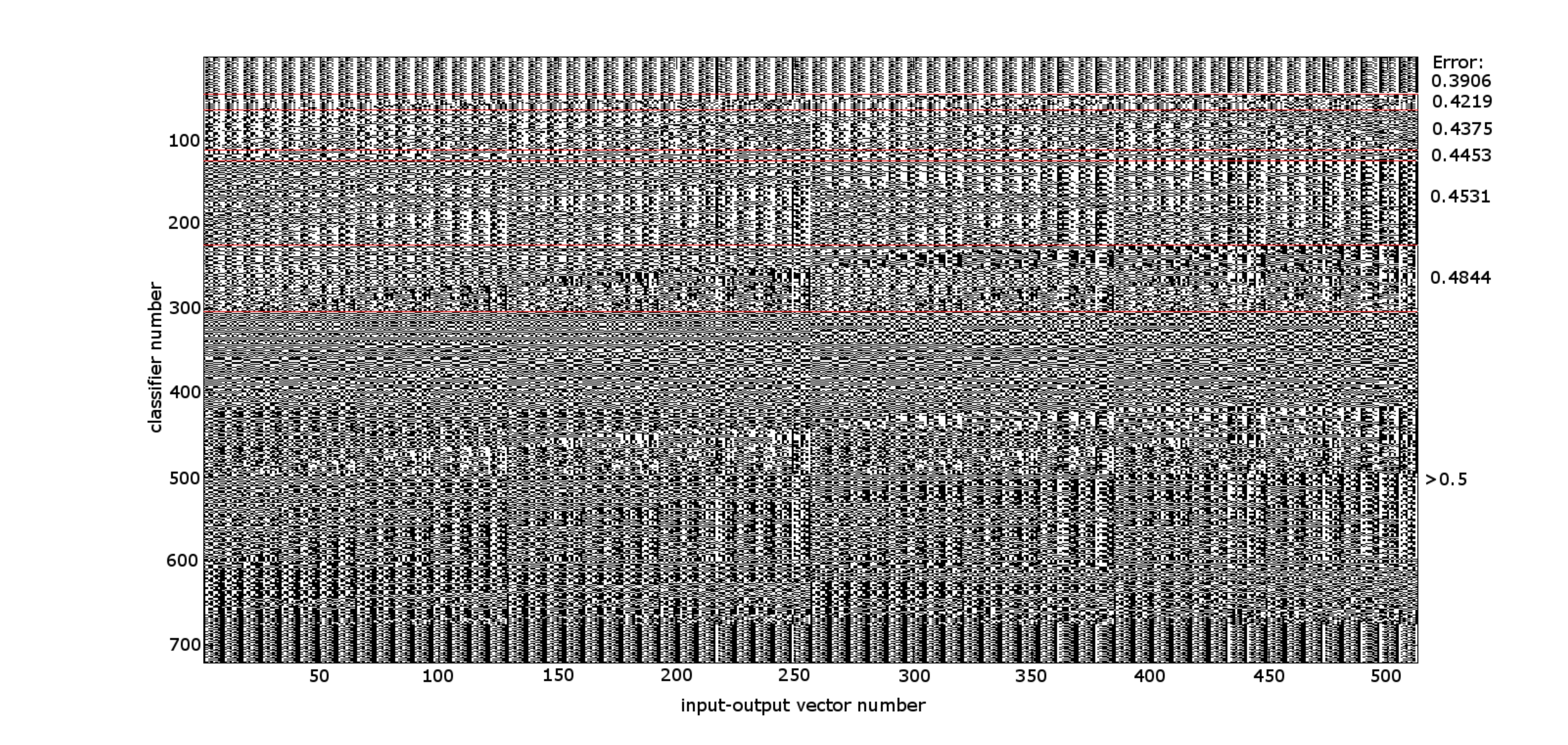}
\caption{Accuracy of weak classifier dictionary on input-output vector space. White/black pixels represent a weak classifier {$h_i(\vec{x})$ (all weak classifiers meeting Condition \ref{cond1} indexed in order of increasing error $\eta_j$ as in Eq. \eqref{eta_j} on vertical axis)} categorizing an 
input-output vector (indexed in lexicographical order on horizontal axis, there are $2^9$ vectors arising from the $9$ Boolean variables in the sample problem) correctly/incorrectly, respectively. 
These classifications were to determine whether an input-output pair was correct or erroneous, i.e., we are analyzing the performance of the specification classifier. 
}
\label{dict_graph}
\end{figure*}

For $Q=16$ through $Q=23$, the error fraction shown is for the best-performing 
classifier, selected from $26$ iterations of the algorithm that were calculated 
using different values of $\lambda$. The consistently observed lack of dependence on 
$\lambda$ in these and other simulations (such as the $50$-iteration result presented 
above) justifies this choice. For $Q=24$ to $Q=26$, it was too computationally 
intensive to run the algorithm multiple times, even on a high performance computing cluster, so the values plotted are from 
a single iteration with $\lambda$ assigned to zero. This was still deemed to be useful 
data given the uniformity of the rest of the simulation results with respect to 
$\lambda$. The dependence on the parity of the number of qubits  
is a result of the potential for the strong classifier to 
return $0$ when the number of weak classifiers in the majority vote is even. Zero is not 
technically a misclassification in that the classifier places the vector $\vec{x}$ in the wrong class, but neither does the classifier give the correct class for $\vec{x}$. Rather, we obtain a \textquotedblleft don't-know\textquotedblright answer from the classifier, which we do not group wtih the misclassifications because it is not an outright error in classification. It is a different, less conclusive piece of information about the proper classification of $\vec{x}$ which may in fact be useful for other applications of such classifiers.

The important conclusion to be drawn from the data quantifying strong classifier errors 
as a function of the number of available qubits is that performance seems to be improving 
only slightly as the number of available qubits increases. This may indicate that 
even with only $16$ qubits, if the algorithm is iterated a sufficient number of times 
to compensate for its random nature, the accuracy achieved is close to the limit 
of what can be done with the current set of weak classifiers. This is encouraging 
in the context of strong classifier generation and sets a challenge for improving the 
performance of weak classifiers or breaking the problem into intermediate stages.

\subsection{Comparison of results with theory}

In light of the conditions for an ideal strong classifier developed in Section 
\ref{conditions}, it is reasonable to ask the following questions: 
How close do the weak classifiers we have for the problem 
studied here come to satisfying
the conditions? What sort of accuracy can we expect our simulations to yield? Fig. 
\ref{dict_graph} and a few related calculations shed some light on the answers. In 
the figure, each row of pixels represents a single weak classifier in the dictionary and each column represents one vector in the input-output space. Horizontal red 
lines divide the different levels of performance exhibited by the weak classifiers. 
White pixels represent a given weak classifier categorizing a given input-output vector 
correctly. Black pixels represent incorrect classifications.

The problematic aspect of Fig. \ref{dict_graph} is the vertical bars of white and black 
exhibited by some of the more accurate classifiers. The method detailed above for 
constructing a completely accurate strong classifier relies on pairs of classifiers which 
are correct where others fall short, and which do not both classify the same input-output 
vector incorrectly. This is impossible to find in the most accurate group of weak 
classifiers alone, given that there are black bars of erroneous classifications spanning 
the entire height of the set.

For numerical analysis of the performance of the set of Boolean weak classifiers on the sample problem, we relate the statistics of the dictionary on the input-output vector space $\mathcal{V}$ to Conditions \ref{cond2} and \ref{cond2a}. Three quantities will be useful for this analysis. The first is the error fraction of an individual weak classifier

\begin{equation}
\label{eta_j}
\eta_j=1-\frac{1}{S}\sum_{s=1}^{S}H\left[y_sh_j(\vec{x_s})\right] ,
\end{equation}
that is, the fraction of the training set incorrectly classified by the weak classifier $h_j(\vec{x})$. We use the Heaviside step function to count the number of vectors correctly classified.

Next is the \emph{minimum possible} overlap of correctly classified vectors for a pair of weak classifiers over $\mathcal{V}$:

\begin{equation}
\label{phi_jj'}
\phi_{jj'}=1-\eta_j-\eta_{j'}
\end{equation}
In Eq. \eqref{phi_jj'}, we add the correctness fraction ($1-\eta_j$) of each weak classifier, then subtract $1$ to arrive at the number of vectors that must be classified correctly by both weak classifiers at once.

The next definition we shall require is that of the \emph{actual} overlap of correct classifications: 

\begin{equation}
\label{gamma_jj'}
\gamma_{jj'}=\frac{1}{S}\sum_{s=1}^{S}H\left[y_s\left(h_j(\vec{x_s})+h_{j'}(\vec{x_s})\right)\right]\equiv\phi_{jj'}+\epsilon_{jj'}
\end{equation}
In Eq. \eqref{gamma_jj'}, we count the number of vectors that are actually classified correctly by both weak classifiers.

If the minimum possible and actual overlaps are the same, i.e., $\epsilon_{jj'}=0$, then Condition \ref{cond2} holds, and the weak classifier pair has minimum \textit{correctness overlap}. Otherwise, if $\phi_{jj'}\ne\gamma_{jj'}$, only the weaker Condition \ref{cond2a} is satisfied, so the weak classifier pair has a greater than minimal correctness overlap and a forced overlap of incorrect classifications $\epsilon_{jj'}>0$ (see Fig. \ref{cond_2_fig}) that could cancel out the correct votes of a different weak classifier pair and cause the strong classifier to be either incorrect or inconclusive.

Our numerical analysis of the weak classifiers satisfying Condition \ref{cond1} (having $\eta_j<0.5$)
showed that the average correctness overlap $\gamma_{jj'}$ between any two weak 
classifiers was $0.3194$. 
The maximum correctness overlap for any pair of weak classifiers was $\gamma_{jj'}=0.6094$. The minimum was $\gamma_{jj'}=0.1563$, between two weak 
classifiers with respective error fractions (amount of the training set misclassified by each individual weak classifier) of $\eta_j=0.4844$ and $\eta_{j'}=0.4531$. Compare this to the 
minimum possible overlap 
with two such classifiers, $\phi_{jj'}=0.0625$, and it becomes apparent that this set of weak classifiers falls short of ideal, given that $\epsilon_{jj'}=0.0938$
for the weak classifier pair with minimum overlap.

When only the most accurate weak classifiers ($\eta_j=0.3906$; above the top red horizontal line in Fig. \ref{dict_graph}) were included, 
the average correctness overlap was $\gamma_{jj'}=0.4389$, the maximum was $\gamma_{jj'}=0.6094$, and the minimum was $\gamma_{jj'}=0.3594$. In order to come up with a generous estimate for the accuracy 
achievable with this group of weak classifiers, we focus on the minimum observed correctness overlap.
The minimum possible correctness overlap for two classifiers with $\eta_j=0.3906$ is $\phi_{jj'}=0.2188$. With an 
ideal set of weak classifiers of error $\eta_{j}=0.3906$ and correctness overlap $\phi_{jj'}=0.2188$, it 
would take seven weak classifiers to construct a completely accurate strong classifier: 
three pairs of two classifiers each to cover a fraction $0.6564$ of the solution space 
with a correctness overlap from one of the pairs, and one more weak classifier to provide 
the extra correct vote on the remaining $0.3436$ fraction of the space. Assuming that 
three pairs of weak classifiers with minimum overlap and optimal relationships to the 
other weak classifier pairs could be found, there will still be a significant error 
due to the overlap fractions of the pairs being larger than ideal. In fact, each pair 
of weak classifiers yields an error contribution of $\epsilon_{jj'}=0.1406$, guaranteeing 
that a fraction $3\epsilon_{jj'}=0.4218$ of the input-output vectors will be classified 
incorrectly by the resulting strong classifier. This is not far from the 
simulation results for odd-qubit strong classifiers (Fig.~\ref{qubits_graph}, left), which suggests
that the algorithm currently in use is producing near-optimal results for the dictionary 
of weak classifiers it has access to.

\section{Conclusions}
\label{conclusions}

We have developed a quantum adiabatic machine learning approach and applied it to the problem of training a quantum software error classifier.
We have also shown how to use this classifier in quantum-parallel on the space of
all possible input-output pairs of a given implemented software program $P$. The training procedure
involves selecting a set of weak classifiers, which are linearly combined,
with binary weights, into two strong classifiers. 

The first quantum aspect
of our approach is an adiabatic quantum algorithm
which finds the optimal set of binary weights as the ground state of a certain
Hamiltonian. We presented two alternatives for this algorithm. The first, 
inspired by \cite{Neven,Neven1}, gives weight to single weak classifiers 
to find an optimal set. The second algorithm for weak classifier 
selection chooses pairs of weak classifiers to form the optimal set and is 
based on a set of sufficient conditions for a completely accurate strong 
classifier that we have developed.

The second quantum aspect of our approach is an explicit
procedure for using the optimal strong classifiers in order to search the
entire space of input-output pairs in quantum-parallel for the existence of an
error in $P$. Such an error is identified by performing an adiabatic quantum
evolution, whose manifold of low-energy final states favors erroneous states. 

A possible improvement of our approach involves adding intermediate
training spaces, which track intermediate program execution states. This has
the potential to fine-tune the weak classifiers, and overcome a limitation
imposed by the desire to restrict our Hamiltonians to low-order 
interactions, yet still 
account for high-order correlations between bits in the input-output
states. 

An additional improvement involves finding optimal interpolation paths
$s(t)$~(\ref{interp}) from
the initial to the final Hamiltonian \cite{PhysRevLett.103.080502,RALZ:10}, for both
the classifier training and classifier implementation problems.

We have applied our quantum adiabatic machine learning approach
to a problem with real-world applications in flight control systems, which 
has facilitated both algorithmic development and characterization of the success 
of training strong classifiers using a set of weak classifiers involving 
minimal bit correlations.


\end{document}